\documentclass[reprint,
superscriptaddress,
 amsmath,amssymb,
 aps,
pra,
floatfix,
]{revtex4-2}
\usepackage{physics}
\usepackage{graphicx}
\graphicspath{ {images/}, {figs/} }
\usepackage{dcolumn}
\usepackage{bm}
\usepackage[mathlines]{lineno}
\usepackage{amssymb}
\usepackage[toc,page]{appendix}
\usepackage{booktabs}
\usepackage{tabularx}
\usepackage{xcolor}
\usepackage{float}
\usepackage{array}
\usepackage{comment}

\usepackage[caption=false]{subfig}
\makeatletter
\newcommand{\colorcaption}[2][]{%
  \begingroup%
  \renewcommand{\@caption@fignum@sep}{ (color online). }%
  \caption[#1]{#2}%
  \endgroup%
}
\makeatother

\makeatletter
\renewcommand{\maketag@@@}[1]{\hbox{\m@th\normalsize\normalfont#1}}%
\makeatother

\usepackage{mathtools}
\usepackage{empheq}
\usepackage{refcount}

\usepackage{multirow}

\usepackage{hyperref}
\hypersetup{
	colorlinks,
	linkcolor={red!50!black},
	citecolor={green!50!black},
	urlcolor={blue!80!black}
}

\definecolor{DarkGreen}{RGB}{0,100,0}

\begin{document}
\title{Non-universal Thermal Hall Responses in Fractional Quantum Hall Droplets}

\author{Fei Tan}
\affiliation{School of Physical and Mathematical Sciences, Nanyang Technological University, 639798 Singapore}

\author{Yuzhu Wang}
\email{yuzhu.wang@ntu.edu.sg}
\affiliation{School of Physical and Mathematical Sciences, Nanyang Technological University, 639798 Singapore}

\author{Xinghao Wang}
\affiliation{International Center for Quantum Materials, School of Physics, Peking University, Beijing 100871, China}

\author{Bo Yang}
\email{yang.bo@ntu.edu.sg}
\affiliation{School of Physical and Mathematical Sciences, Nanyang Technological University, 639798 Singapore}
\affiliation{Institute of High Performance Computing, A*STAR, 138632 Singapore}
\date{\today}

\begin{abstract}
We analytically compute the thermal Hall conductance (THC) of fractional quantum Hall droplets under realistic conditions that go beyond the idealized linear edge theory with conformal symmetry. Specifically, we consider finite-size effects at low temperature, nonzero self-energies of quasiholes, and general edge dispersions. We derive measurable corrections in THC that are consistent with the experimental observables. Although the quantized THC is commonly regarded as a topological invariant that is independent of edge confinement, our results show that this quantization remains robust only for arbitrary edge dispersion in the thermodynamic limit. Furthermore,  the THC contributed by Abelian modes can become extremely sensitive to finite-size effects and irregular confining potentials in any realistic experimental system. In contrast, non-Abelian modes show robust THC signatures under perturbations, indicating an intrinsic stability of non-Abelian anyons.
\end{abstract}

\maketitle

\textit{Introduction.--} In a two-dimensional electron gas (2DEG) under strong magnetic fields and low temperatures, the Hall resistance exhibits plateaus at rational values of $h/e^2$, which is known as the fractional quantum Hall (FQH) effect \cite{klitzing1980new, tsui1982two}. Such quantization cannot be explained by single-particle physics but arises from a topological ground state stabilized by quenched kinetic energy and broken time-reversal symmetry \cite{laughlin1981quantized, halperin1982quantized, moore1991nonabelions}. The quasiparticles of FQH fluids carry fractional charge and obey anyonic statistics \cite{laughlin1983anomalous, arovas1984fractional, stern2008anyons, halperin1984statistics}, and in certain non-Abelian phases they realize fault-tolerant topological qubits \cite{das2005topologically, nayak2008non, stern2013topological,sarma2015majorana}. Analogous lattice realizations called fractional Chern insulators have been observed in twisted bilayer materials and multilayer graphene \cite{cai2023signatures, park2023observation, han2024large}. 

In topological orders like quantum Hall (QH) fluids, there exist quantized responses that are insensitive to local perturbations \cite{avron1983homotopy, berry1984quantal, laughlin1983anomalous, moore1991nonabelions, niu1990ground, li2008entanglement, fukui2012bulk}. Because the bulk is gapped, transport signatures originate from the gapless chiral edge modes that encode the same topological data \cite{halperin1982quantized,hatsugai1993chern,hatsugai1993edge}. Besides the quantized electrical Hall conductance, the thermal Hall conductance (THC) provides an independent probe of edge central charge and thus of the underlying topological order \cite{kane1997quantized}. Recent experiments have shown that the quantized value of the THC could be used to distinguish different candidate phases, especially at the half-filling \cite{banerjee2018observation, paul2024topological, dutta2022isolated,dutta2022distinguishing, srivastav2022determination}.

The edges of QH phases are effectively described by chiral Luttinger liquids \cite{wen1990chiral, haldane1981luttinger}. The THC from chiral edge modes is proportional to the central charge $c$ of the underlying $(1+1)$D conformal field theory (CFT) that characterizes the edge \cite{affleck1988universal, kane1997quantized,  cappelli2002thermal,bernevig2008properties,fukusumi2024edge,fukusumi2025gauging}. The universality of the THC thus depends on conformal symmetry, typically requiring a linear dispersion in the edge modes. However, under realistic experimental conditions, additional factors can contribute to deviations in the measured THC, such as insufficient thermal equilibration due to finite-size effects, quasihole self-energy corrections induced by non-ideal interactions, and nontrivial edge dispersions arising from irregular confining potentials. Considerations of realistic conditions bring non-universal correction terms to the universal quantized value. These effects are especially relevant for understanding why thermal Hall measurements have so far shown significantly lower precision than their electric counterparts, and the recently observed THC value at half-filling with a controversial origin \cite{banerjee2018observation, paul2024topological, dutta2022isolated, dutta2022distinguishing, srivastav2022determination,simon2018interpretation,feldman2018comment,simon2018reply,roy2025half}.

In this Letter, we develop a microscopic framework to analyze deviations of the THC $\kappa$ in FQH states arising from finite size, quasihole self-energies, and nonlinear edge dispersions. Rather than relying on mesoscopic transport modeling such as the Landauer-Buttiker formalism, we construct edge partition functions for both Abelian and non-Abelian phases, whose ground states are Jack polynomials \cite{bernevig2008model,bernevig2008generalized}, and derive $\kappa$ from thermodynamic relations. Numerical calculations confirm that $\kappa$ of non-Abelian edges remains quantized in the canonical ensemble, while Abelian modes exhibit strong finite-size corrections. Nonlinear dispersions further break the universal relation between specific heat, $\kappa$, and central charge. These non-universal corrections is consistent with the recent experimental deviations \cite{banerjee2017observed,banerjee2018observation,melcer2022absent} and can help distinguish competing $\nu=5/2$ candidate phases.

\textit{Thermal Hall conductance on a disk.--} A temperature gradient across a Hall bar induces a transverse heat current $J_Q$, known as the Leduc–Righi effect, which is the thermal analog of the Hall effect \cite{leduc1888modifications,onose2010observation,pitaevskii2012physical}. The THC, $\kappa=\partial J_Q/\partial T$, is predicted to be quantized as \cite{kane1997quantized}
\begin{equation}
\kappa = \kappa_0 \: T \:c\:, \qquad \kappa_0=\pi^2 k_B^2/(3h),
\end{equation}
where $c$ is the edge CFT central charge and $T$ is the electron temperature. Each chiral boson contributes $c=1$, while a Majorana mode contributes $c=1/2$ \cite{friedan1984conformal,moore1991nonabelions,cappelli2002thermal}. Counterpropagating modes contribute with opposite signs, making intermode thermal equilibration crucial for observing quantized $\kappa$ \cite{kane1997quantized}.

\begin{figure}[t]
    \centering
    \includegraphics[width=\columnwidth]{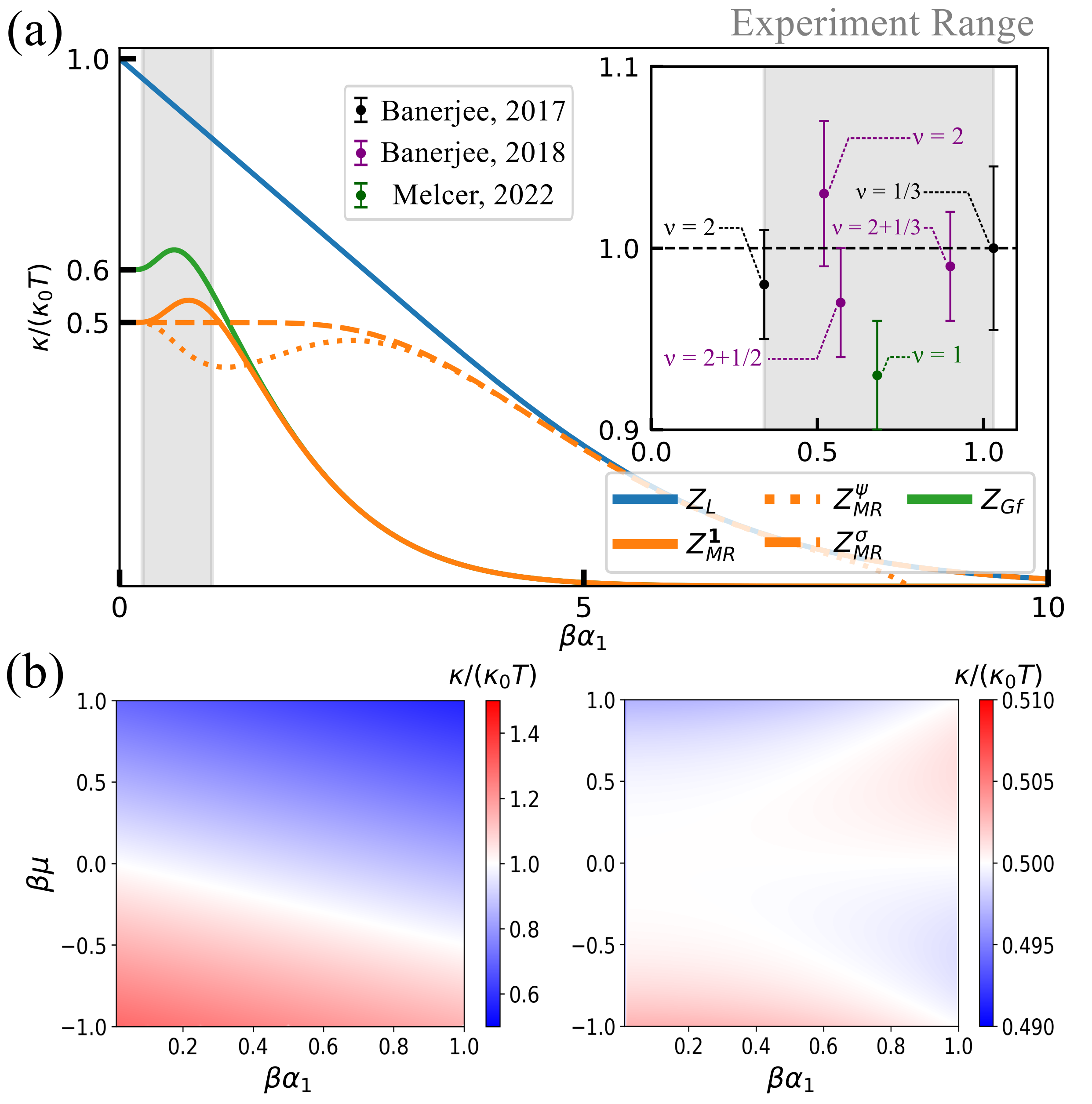}
    \caption{\textbf{Non-universal thermal Hall responses of different edge modes.}(a) Thermal Hall conductance $\kappa/(\kappa_0 T)$ versus $\beta\alpha_1$ for Abelian $U(1)$ (blue), Majorana (orange, different line styles for distinct sectors), and the non-Abelian (NA) component of the Gaffnian (green) edge modes. At $\beta\alpha_1=0$, $\kappa/(\kappa_0 T)$ equals the central charge. The $U(1)$ mode decreases nearly linearly with $\beta\alpha_1$, while the Majorana and NA modes remain almost constant, revealing the robustness of non-Abelian excitations against finite-size effects. The inset shows the experimental regime (gray area), where the error in available $U(1)$ data is proportional to $\alpha \beta$, consistent with theory \cite{banerjee2017observed, banerjee2018observation, melcer2022absent}. (b) Dependence of $\kappa$ on quasihole creation energy and edge-mode velocity: the $U(1)$ mode (left) varies linearly with these parameters, whereas the Majorana mode (right) remains essentially unchanged, underscoring its insensitivity to microscopic details.
}
    \label{fig:finiteTem and Quasihole}
\end{figure}

Let $\Delta k$ denote the minimal momentum spacing of quasihole states relative to the ground state and $v_F$ the Fermi velocity of the corresponding edge mode, assumed constant for each mode. In the thermodynamic regime (high temperature compared with level spacing),
$k_B T \gg \hbar v_F \Delta k$, the THC remains quantized and thus topological. Because $\Delta k$ scales inversely with the system size $L$, the quantization depends on the competition between temperature $T$ and $L$ (The length scale is set by the thermal equilibration length of edge modes). To characterize this interplay, we introduce the dimensionless parameter $\beta\alpha_1$, where $\alpha_1 = hv_F/L$ (for a linear dispersion) and $\beta = 1/(k_B T)$.

To understand edge excitations microscopically, we consider an FQH fluid on a disk with circumference $L$ and $N$ electrons. Due to rotational invariance, angular momentum $m$ serves as a good quantum number for classifying states. When a magnetic flux is adiabatically inserted into the bulk, electrons will get pushed towards the edge by a radial current, effectively creating a quasihole in the bulk \cite{laughlin1981quantized}. Each quasihole induces a density modulation at the edge and thus a specific edge mode. Edge excitations can therefore be classified by the angular momentum of the corresponding quasihole states: if $m_0$ is the ground-state angular momentum, we denote by $p(\Delta m)$ the degeneracy in the sector $m=m_0+\Delta m$. In the thermodynamic limit, the edge forms a one-dimensional channel with linear momentum $k=m/L$. This is a working example of the bulk–edge correspondence \cite{wen1992theory,wen1994chiral,yan2019bulk}.

\textit{Finite-size corrections to THC.--}
In experiments, the temperature cannot be freely tuned, as it must remain low enough to preserve the FQH phase. We therefore refer to the corrections associated with a finite $\alpha_1$ as finite-size effects. Physically, this corresponds to regimes where the system size or temperature is insufficient for reaching the thermodynamic limit.

We first consider the chiral $U(1)$ bosonic edge mode common to all Laughlin states, including the integer quantum Hall phases. As a concrete example, we take the $\nu = 1/3$ Laughlin state with linear dispersion and vanishing quasihole self-energies. For any Laughlin state at filling $\nu = 1/(2k+1)$ with $k \in \mathbb{Z}+$, the degeneracy of quasihole states at angular-momentum sector $\Delta m$ follows the Virasoro counting identical to the integer partition number $p(\Delta m)$, i.e., the number of ways positive integers sum to $\Delta m$ \cite{wen1994chiral}. This yields the partition function for Abelian edge excitations in the thermodynamic limit ($N \to \infty$):
\begin{equation}
Z_L^{(\infty)} = \sum_{\Delta m=0}^{\infty} p(\Delta m) \cdot e^{-\beta \alpha_1 \Delta m}
= \frac{1}{(q)_\infty},
\label{laughlin_Z}
\end{equation}
where $q = e^{-\beta \alpha_1}$, $(q)_n = \prod_{i=1}^n (1-q^i)$, and $(q)_\infty = \lim_{n\to\infty} (q)_n$. The ground-state contribution $\propto q^{N^2/2}$ has been omitted \cite{bernevig2008properties}. Mathematically, Eq.~\ref{laughlin_Z} is the generating function for unrestricted integer partitions \cite{andrews1998theory}. In the asymptotic limit $\beta \alpha_1 \to 0$, the specific heat of the Laughlin edge is \footnote{\label{note:appendix}Further details are provided in the Appendix.} $C_{L} \sim C_0 - \tfrac{1}{2} + \mathcal{O}(\beta,\alpha_1)$, where $C_0 = \pi^2/(3\beta\alpha_1)$ and “$\sim$” denotes asymptotic equivalence. The corresponding thermal Hall conductance under linear dispersion is then
\begin{equation}
\kappa_L = \frac{v_F}{L} C_{L}
= \kappa_0 T \left( 1 - \frac{3}{2\pi^2}\beta \alpha_1 + \mathcal{O}(\beta^2,\alpha_1^2) \right),
\end{equation}
which yields a central charge $c=1$ for the Laughlin edge in the limit $\beta \alpha_1 \to 0$.

Using a Fermi velocity $v_F \approx 10^5$ m/s for the chiral boson edge state in GaAs 2DEGs \cite{roosli2020observation,sahasrabudhe2018optimization}, the experimental value of $\beta \alpha_1$ lies between $0.3$ and $1.0$ as shown in Fig.~\ref{fig:finiteTem and Quasihole}a. In transport measurements, a grand canonical treatment may provide more accurate estimates of $\kappa$ under strong finite-size effects \cite{cappelli2002thermal}. The measured THC in the $\nu=1/3$ Laughlin phase was $\kappa = (1.0 \pm 0.045)\:\kappa_0 T$ \cite{banerjee2017observed}, with an uncertainty on the order of $10^{-2}$, significantly larger than that of charge Hall conductance measurements.

We now study the thermal Hall response in non-Abelian phases, focusing on the half-filled Moore–Read (MR) state. Two types of modes contribute to its edge excitations: the Abelian chiral boson modes and the non-Abelian Majorana fermion modes \cite{moore1991nonabelions,milovanovic1996edge}. The former corresponds to the $U(1)_2$ sector, while the latter is described by the Ising CFT with primary fields ${\mathbf{1},\psi,\sigma}$ \cite{francesco2012conformal}. In the thermodynamic limit, MR quasihole states satisfy the generalized Pauli principle: no more than two particles may occupy four consecutive orbitals, stemming from the model Hamiltonian constraint \cite{bernevig2008model}. The counting of MR quasihole states in successive angular-momentum sectors, $p_p(\Delta m)$, follows $1,1,3,5,10,16,\ldots$, yielding the partition function in the $N \to \infty$ limit \cite{bernevig2008properties,andrews1998theory}:
\begin{equation}
     Z_{\text{MR}}^{(\infty)} 
     = \frac{1}{(q)_{\infty}} \sum_{n=0}^{\infty}\frac{q^{\frac{n^2}{2}}}{(q)_n} = \frac{1}{(q)_\infty} \prod_{j=0}^{\infty} \left(1 + q^{j + 1/2} \right).
\label{mr}
\end{equation}
The factor $1/{(q)_{\infty}}$ in Eq.~\ref{mr} represents the Abelian $U(1)_2$ charge sector (as in Eq.~\ref{laughlin_Z}), while the remaining piece encodes the neutral Majorana fermion modes. Note that Eq.~\ref{mr} sums over the two partition functions (i.e., the Neveu-Schwarz (NS) characters) $Z_{\text{MR}}^{\mathbf{1}}$ and $ Z_{\text{MR}}^{\mathbf{\psi}}$ \cite{milovanovic1996edge}, and thus does not resolve parity subsectors. In a finite droplet, however, such a distinction is essential since locality of the electron operator requires a \textit{gluing condition} between the Majorana and $U(1)_2$ sectors~\cite{ino1999multiple,milovanovic1996edge, sohal2020entanglement}. Microscopically, this means that the parity of the electron number and the distribution of bulk anyons fix the Majorana sector. Hence on a disk, even $N$ selects $Z_{\text{MR}}^{\mathbf{1}} $, while odd $N$ selects $Z_{\text{MR}}^{\mathbf{\psi}}$ \footnotemark[\value{footnote}].

In contrast to the NS sectors, the Ramond (R) sector $Z_{\text{MR}}^{\mathbf{\sigma}} $ does not contribute to the edge partition function unless bulk $-e/4$ quasiholes are present, as the fusion rules $\psi \times \sigma = \sigma$ and $\sigma \times \sigma=1+\psi$ suggest. One way to visualize this is by the Wilson-line picture on a cylinder, where the line connecting an anyon pair enforces the parity constraint, equivalent to the bulk-edge correspondence upon mapping to a disk \footnotemark[\value{footnote}]. The presence or absence of these bulk quasiholes gives rise to the well-known \textit{odd-even effect} of the Moore-Read state as a direct signature of non-Abelian statistics, i.e., the interference pattern in a $\nu=5/2$ Fabry-P\'erot interferometer is predicted to depend on whether the number of bulk $-e/4$ quasiholes is even or odd, with the odd case suppressing interference \cite{stern06proposed, Bonderson06detecting}. 

In the thermodynamic limit, both the sector- and parity-resolved Moore–Read (MR) partition functions approach the same chiral central charge $c=3/2$ and thus the same quantized THC \footnotemark[\value{footnote}]. In the asymptotic limit $\beta\alpha_1 \to 0$, the Majorana contribution to the specific heat is $C_{MF}\sim C_0/2+\mathcal{O}(\beta^2,\alpha_1^2)$, with the linear correction in $\beta\alpha_1$ vanishing for all parity and boundary conditions—signaling the intrinsic robustness of non-Abelian modes. As shown in Fig.~\ref{fig:finiteTem and Quasihole}a, this robustness persists near $\beta\alpha_1 = 0$, while larger deviations render $\kappa$ parity- and sector-dependent. Such parity-resolved shifts offer an additional probe of non-Abelian order. The total THC for the MR edge combining $U(1)_2$ bosons and Majorana fermions is
\begin{equation}
\kappa_{MR}=\kappa_0 T \left(\frac{3}{2}-\frac{3}{2\pi^2}\beta\alpha_1+\mathcal{O}(\beta^2,\alpha_1^2)\right),
\end{equation}
confirming $c=1+\tfrac{1}{2}=\tfrac{3}{2}$, with finite-size corrections dominated by the Abelian mode.

\textit{Non-zero self-energy of quasiholes.--} Quasiholes and thus edge modes are conventionally treated as a non-interacting ``ideal gas'', which serves as a good approximation in the dilute limit. Tuning on the interactions between edge modes generally renormalizes the Fermi velocity $v_F$ without significantly affecting THC \cite{kane1997quantized,hu2009edge}. Additionally, quasiholes are assumed to be ``massless", meaning their creation does not require finite energy upon flux insertion. However, this condition is generally not true with realistic interactions between electrons, as confirmed by extensive numerical calculations \cite{wojs2001interaction, xu2025dynamics}. To capture such effects, we write the partition function as:
\begin{equation}
\begin{aligned}
    Z_{qh}&=\sum_{\Delta m=0}^{\infty} \Tilde{p}(\beta, \Delta m) \cdot e^{-\beta \alpha_1 \Delta m},
\end{aligned}
\end{equation}
where the effective density of state at a finite temperature $\Tilde{p}(\beta, \Delta m)$ reads: 
\begin{equation}
\begin{aligned}
    \Tilde{p}(\beta, \Delta m) &= \sum_{\xi=1}^{p(\Delta m)} e^{-\beta \Tilde{\epsilon}_{m, \xi}}.
\end{aligned}
\end{equation}
Here $\Tilde{\epsilon}_{m, \xi}$ depends on the details of the quasihole states labeled by $\xi$ within the angular momentum $m$ sector, which generically contains a different number of interacting quasiholes. Since increasing temperature eventually destroys FQH phases, the original density of states $p(\Delta m)$ will reappear only when the quasihole self-energy is small compared to the thermal energy. This also implies that conformal invariance is effectively restored at the zero self-energy limit.

We now reconsider the Laughlin phase. When a quasihole is created in the FQH droplet, the total energy of the quantum fluid decreases due to the repulsive interactions among electrons. As a result, quasiholes acquire a negative self-energy (or “mass”). Assuming that each flux insertion carries a constant energy cost $\mu$, and that the quasiholes form a dilute, non-interacting gas, we can write down the modified partition function of the Laughlin edge modes as:
\begin{equation}
\begin{aligned}
    Z_{L,qh}&=\sum_{\Delta m=0}^{\infty} \sum_{\xi=1}^{p(\Delta m)} e^{-\beta \alpha_1 \Delta m} e^{-\beta \tilde{\epsilon}_{m, \xi}}
    = \prod_{i=1}^{\infty} \frac{1}{1- tq^i},
\label{eqn:Laughlin self energy}
\end{aligned}
\end{equation}
where $t\equiv e^{-\beta \mu}$, and $\Tilde{\epsilon}_{m, \xi} = \epsilon_{m, \xi} -\alpha_1\Delta m$ is the total quasihole self-energy of the excitation state $\xi$, i.e, the product of $\mu$ and the number of quasiholes in state $\xi$. If we further assume the velocity of the edge mode remains the same as in the ideal case, the asymptotic THC is now:
\begin{equation}
    \kappa_{L, qh} = \kappa_0 T \bigg(1 - \frac{3}{2\pi^2}\beta (\alpha_1 + 2 \mu) +\mathcal{O}(\beta^2,\alpha_1^2,\mu^2)\bigg)
\end{equation}
which agrees well with numerical calculations in Fig.~\ref{fig:finiteTem and Quasihole}b. The additional correction enhances the THC when the quasihole creation energy $\mu$ is negative, since it increases the effective density of states at finite temperatures.

Similarly, we can obtain the modified THC of the MR phase. Assuming that all types of quasiholes contributing to both chiral boson mode and Majorana fermion mode have the self-energy $\mu$, the partition function of the Majorana fermion mode is given by:
\begin{equation}
    Z_{MF, qh} = \prod_{n=0}^{\infty} (1+q^{n+\frac{1}{2}}t^{\frac{1}{2}}),
\end{equation}
while the resultant THC turns out to be invariant:
\begin{equation}
\begin{aligned}
    \kappa_{MF, qh} =\kappa_0T \left( \frac{1}{2} + \mathcal{O}(\beta^3,\alpha_1^3,\mu^2) \right)
    .
\end{aligned}
\end{equation}
in the sense that the THC does not linearly dependent on $\beta \alpha_1$ and $\beta \mu$.

\begin{figure*}[t]
    \centering
    \includegraphics[width=\linewidth]{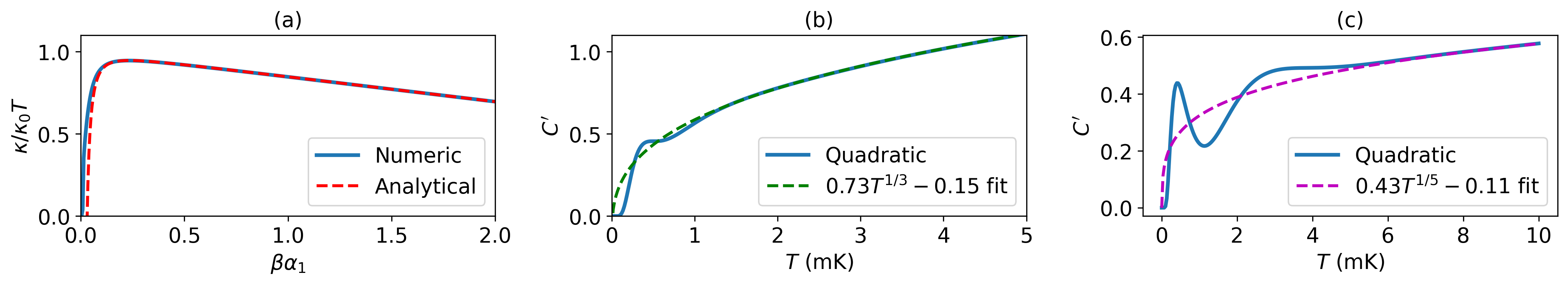} 
    \caption{\textbf{Thermodynamic observables under nonlinear dispersions.} (a) Thermal Hall conductance (THC) of Abelian modes for $\epsilon_m=\alpha_1\Delta m+\alpha_2(\Delta m)^2$ with $\alpha_2=10^{-4} \ll \alpha_1$. The dashed line shows the analytical result from Eq.~\ref{eqn:1+2} in the limit $\beta\alpha_1\to 0$. (b,c) Normalized specific heat $C'=(k_B/\alpha_n^2)C_L$ versus temperature under quadratic and cubic dispersions, with $\alpha_2,\alpha_3 \sim 10^{-26}$. The dashed lines indicate high-temperature fits. For quadratic (cubic) dispersions, $C_L \propto T^{1/3}$ ($T^{1/5}$).} 
    \label{fig:QuadraticHeat}
\end{figure*}

\textit{Generic THC with nonlinear dispersion.--} In the thermodynamic limit, our microscopic approach with discrete angular momentum $m$ becomes equivalent to the Luttinger liquid formalism with continuous momentum $k$ \cite{luttinger1960fermi, luttinger1963exactly, haldane1981luttinger, kane1997quantized,haldane2005luttinger}. Taking the chiral $U(1)$ boson modes as an example, we can establish the correspondence by invoking $f_B(\epsilon_k)=(e^{\beta \epsilon_k}-1)^{-1}$ with an arbitrary dispersion relation $\epsilon_k$ and derive the THC as:
\begin{equation}
    \kappa_{U(1)} = \frac{1}{2\pi} \int_{0}^{\infty} \: dk \: \frac{\partial \epsilon_k}{\partial k} \: \frac{(\beta \epsilon_k)^2 \: e^{\beta \epsilon_k}}{(e^{\beta \epsilon_k}-1)^2} = \frac{\pi^2 k_B^2}{3h} T.
\label{eqn:LinearTHC}
\end{equation}
which is irrelevant to $\epsilon_k$ and thus universal. However, when finite-size effects are taken into account, such universality vanishes, which can be seen from the Euler-Maclaurin formula \footnotemark[\value{footnote}]. Moreover, the commonly assumed linear dispersion of edge modes relies on the idealization that the confining potential at the sample boundary is perfectly quadratic. In realistic QH systems, the edge potential is generally not strictly quadratic, and deviations from this assumption can lead to significant nonlinearities in the edge-mode dispersion. It is therefore instructive to analyze simplified models that go beyond the linear-dispersion approximation, where the conformal symmetry of the edge theory is explicitly broken.

We first consider the case where the energy dispersion is not perfectly linear $\epsilon_m = \alpha_1 \Delta m + \alpha_2(\Delta m)^2$, where $\alpha_2 \ll \alpha_1$ is the quadratic dispersion coefficient. By taking the Laughlin state at $\nu=1/3$ and assuming that the degeneracy of each $\Delta m$ sector is not affected, we can obtain the asymptotic THC ($\beta \alpha_1 \rightarrow 0$) as:
\begin{equation}
    \kappa_L \approx \kappa_0 T\left( 1-\frac{3}{2\pi^2}\beta \alpha_1 -\pi^2\frac{\alpha_2}{\alpha_1^3 \beta^2}\right)
\label{eqn:1+2}
\end{equation}
where the extra correction term is due to the quadratic component of the dispersion relation \footnotemark[\value{footnote}]. This is further confirmed by numerical results as shown in Fig.\ref{fig:QuadraticHeat}(a), in which the THC is reduced and the quantity $\kappa/(\kappa_0T)$ at the limit of $\beta \alpha_1 \rightarrow 0$ is no longer representing the central charge.

Next, we show that the specific heat will be decoupled from the THC if the energy dispersion is purely nonlinear. We consider the partition function for $\nu=1/3$ Laughlin state (as in Eqn.\ref{laughlin_Z}) but with the power-law dispersion $\epsilon_{m}=\alpha_n (\Delta m)^n$, where $\alpha_n$ is the dispersion coefficient. In the limit $\beta\alpha_n \to 0$, the partition function yields a power-law relation \footnotemark[\value{footnote}] between specific heat and temperature: 
\begin{equation}
    C_L \propto T^{\frac{1}{2n-1}},
\end{equation}
which is consistent with numerical results in Fig.~\ref{fig:QuadraticHeat}. When the dispersion becomes nonlinear, however, the THC no longer scales linearly with the heat capacity because the edge-mode velocity varies with angular momentum $m$. For a general FQH edge with degeneracy $p(\Delta m)$, the thermal current reads
\begin{equation}
J_Q=\sum_{\Delta m=0}^{\infty} v_m\:\frac{\epsilon_m}{L}\frac{p(\Delta m)e^{-\beta\epsilon_m}}{Z},
\label{eqn:GeneralThermalCurrent}
\end{equation}
giving $\kappa=\partial J_Q/\partial T$. The conventional relation $\kappa=v_FC/L$ holds only for a strictly linear dispersion \footnotemark[\value{footnote}]. Otherwise Eq.~\ref{eqn:GeneralThermalCurrent} must be used, explaining the breakdown of Eq.~\ref{eqn:LinearTHC} in realistic edge spectra.

\textit{Summary and Discussions.--} We have analyzed the nonuniversal behavior of the THC in both Laughlin and Moore–Read states under finite-size, quasihole self-energy, and nonlinear-dispersion effects. The non-Abelian contribution from Majorana modes remains robust against finite-size corrections, indicating intrinsic stability, while the $U(1)$ bosonic mode decreases linearly with $\beta\alpha_1$, consistent with deviations observed in experiments. Finite quasihole energies preserve quantization in the non-Abelian sector, but nonlinear dispersions break the usual proportionality between THC, specific heat, and central charge, leading to nonuniversal responses in mesoscopic systems. 

These results offer a framework for interpreting recent measurements of $\kappa\simeq2.5$ that is often attributed to a particle–hole Pfaffian phase despite numerical evidence to the contrary \cite{banerjee2018observation,dutta2022distinguishing}. A full explanation likely requires incorporating edge reconstructions, disorder, unequal mode velocities, incomplete thermal equilibration, and momentum mismatches between counterpropagating branches \cite{simon2018interpretation,lotrivc2025majorana}. Future experiments capable of resolving parity-resolved or energy-dependent deviations in $\kappa$ will be crucial for identifying the true nature of the observed half-quantized thermal Hall plateau.

Our results provide additional experimentally accessible parameters from a different perspective that can help distinguish the contributions of different edge modes. Although all statistical ensembles converge in the thermodynamic limit, notable discrepancies emerge in mesoscopic systems. While a grand canonical ensemble typically describes transport measurements, we propose an alternative calorimetric approach using quantum Hall droplets, which naturally realize a canonical ensemble. Such droplets can be engineered through electrostatic confinement or gate-defined potentials \cite{Cano2013}. By locally injecting power into edge channels—via noise or bias at a point contact—and monitoring the resulting temperature rise, one can extract both the specific heat and the THC. Recent local-power techniques have already demonstrated the feasibility of such measurements even without full thermal equilibration \cite{Melcer2023,LeBreton2022}. Looking ahead, moir\'e platforms hosting FCIs, such as twisted MoTe$_2$, provide ideal testbeds since their micron-scale flakes feature low heat capacity, sharp edges, and van der Waals interfaces compatible with contactless thermometry. These advances may soon enable direct calorimetric probes of $\kappa$ in a canonical setting.

\begin{acknowledgments}
F. Tan and Y. Wang contribute equally to this work. We wish to thank F. D. M. Haldane for the inspiring discussion on the Luttinger liquid approach to the universal THC values, S. H. Simon for the comments on the non-Abelian case, and Y. Fukusumi for the fruitful discussions on CFT and semion representations. We thank D. T. Son, A. Sandvik, X. Yang, Ha Q. Trung, Q. Xu and G. Ji for the useful discussions. This work is supported by the NTU grant for the National Research Foundation, Singapore under the NRF fellowship award (NRF-NRFF12-2020-005), and Singapore Ministry of Education (MOE) Academic Research Fund Tier 3 Grant (No. MOE-MOET32023-0003) “Quantum Geometric Advantage.”
\end{acknowledgments}

\bibliographystyle{unsrt}
\bibliography{sample}

\newpage
\appendix
\counterwithin{figure}{section}

\setcounter{table}{0}
\renewcommand{\thetable}{A.\arabic{table}}
\onecolumngrid

\section*{Supplementary material of ``Non-universal Behaviors of Thermal Hall Conductance in Fractional Quantum Hall States"}

In the supplementary material, we provide detailed technical analyses that support and extend the results in the main text. To help readers quickly locate topics of interest, we provide a summary of the content for each section below. In Sec.~\ref{AppendixA}, we develop the Mellin transform method in analytic number theory to study the asymptotic behavior of logarithmic generating functions and thus the finite-size and nonzero-self-energy corrections to the thermal Hall conductance (THC), with detailed discussions for Laughlin phases (chiral boson modes), different sectors in the Majorana fermion phase, and the non-Abelian modes in the Gaffnian phase. In Sec.~\ref{AppendixB}, we show that the THC ceases to be universal under the combined action of finite-size effects and a general dispersion relation. In Sec.~\ref{AppendixC}, we specialize to the Laughlin $\nu = 1/3$ state and examine its THC under the quadratic correction to the dispersion relation ($\epsilon = \alpha_1 n + \alpha_2 n^2$). In Sec.~\ref{AppendixD}, we provide exact derivations of the heat capacity to leading order for arbitrary power-law dispersions of Laughlin edge modes, where $\epsilon_m = \alpha_k (\Delta m)^k$. In Sec.~\ref{AppendixE}, we establish the general relation between the THC (or thermal current) and the partition function $Z$, independent of the underlying dispersion. Finally, in Sec.~\ref{AppendixF}, we compute the experimental range of $\beta \alpha_1$ by using the data from previous experimental work. We also compare the experimental error bars of the THC measured in GaAs systems with the finite-size correction term derived in this work.


\section{Asymptotic behavior of thermal Hall conductance \label{AppendixA}}
In this section, we introduce a powerful technique in analytic number theory called the \textit{Mellin transform}, which can help solve the asymptotic behavior of logarithmic generating functions near the singularity at $0$ or $\infty$ with series-product identities, where it is normally hard to solve the Laurent series directly. In our case, we are interested in knowing the expression of THC when $q\rightarrow0$ (or $\beta \alpha_1 \rightarrow 0$). However, the THC is not well-defined at this point since it will diverge so one has to study the asymptotic behavior of the functions (see \ref{eqnExactC}).

\subsection{Mellin transform}
For $x \in \mathbb{R}_{+}$,  the Mellin transform of a function $f(x)$is defined as:
\begin{equation}
f^{\star}(s) =\int_0^{\infty} f(x) \cdot x^{s-1} d x, \quad s \in \mathbb{C}.
\end{equation}
Here $s$ should be constrained to a strip, i.e. $a< \text{Re}(s) <b$ where $f^*$ exists. The inverse transform can be written as:
\begin{equation}
f(x)=\frac{1}{2 \pi i} \int_{c-i \infty}^{c+i \infty} f^{\star}(s) \cdot x^{-s} d s,
\end{equation}
and we denote the Mellin dual as:
\begin{equation}
f(x) = \mathcal{M}^{-1}[ f^{\star}(s)], \quad f^{\star}(s) = \mathcal{M}[ f(x)].
\end{equation}
This transform has the harmonic sum property:
\begin{equation}
\sum_j \lambda_j f\left(\omega_j \cdot x\right) \stackrel{\mathcal{M}}{\mapsto}\left(\sum_j \lambda_j \cdot \omega_j^{-s}\right) \cdot f^{\star}(s),
\end{equation}
which implies that if one can identity the ``amplitude'' $\lambda_j$ and the ``frequency'' $\omega_j$ from the summation, one can then factorize the harmonic sum of $f(x)$ to the product of a generalized Dirichlet series and $f^*(s)$. Essentially, the asymptotic expansion of different orders at $0$ or $\infty$ is described by the residues at different poles. This is extremely powerful when dealing with generating functions, considering they are essentially formal infinite function series.

Some properties of Mellin transforms and the Mellin dual functions that we will use in this paper are presented in Table.~\ref{table:MellinDual}:

\begin{table}[h]
\centering 
\renewcommand{\arraystretch}{1.6}
\begin{tabular}{c|c|c}
\hline\hline
\textbf{Function} & \textbf{Mellin Transform} & \textbf{Fundamental Strip} \\ \hline
        $f(x)$ & $f^{\star}(s)=\int_0^{\infty} f(x) x^{s-1} dx$ & $a< \text{Re}(s) <b$\\ \hline
        $x^{\nu}f(x)$ & $f^{\star}(s+\nu)$ & $a-\text{Re}(\nu)< \text{Re}(s) <b-\text{Re}(\nu)$ \\ \hline  $f'(x)$ &  $-(s-1)\cdot f^{\star}(s-1) $  &  $a+1< \text{Re}(s) <b+1$\\ \hline 
         $x^{\nu}f(x)$ &  $f^{\star}(s+\nu) $  &  $a+1< \text{Re}(s) <b+1$\\ \hline
        $e^{-px},p>0$ &  $p^{-s} \cdot \Gamma(s) $  &  $0< \text{Re}(s) <\infty$\\ \hline
        $(e^{ax} -1)^{-1},Re(a)>0$ &  $a^{-s} \cdot \Gamma(s) \cdot \zeta(s) $  &  $1< \text{Re}(s) <\infty$\\ \hline
        $(e^{ax} + 1)^{-1},Re(a)>0$ &  $a^{-s} \cdot \Gamma(s) \cdot \zeta(s) \cdot (1-2^{1-s})$  &  $0< \text{Re}(s) <\infty$\\ \hline
        $(e^{-ax})(1-e^{-x})^{-1},Re(a)>0$ &  $\Gamma(s) \cdot \zeta(s,a)$  &  $1< \text{Re}(s) <\infty$\\ \hline\hline
\end{tabular}
\caption{Some special function and their Mellin transformed functions. The third column shows the constraint of $s$ at the strip \cite{flajolet2009analytic}.}
\label{table:MellinDual}
\end{table}

\subsection{Chiral bosonic edge mode}
\begin{figure}
  \includegraphics[width=0.6\linewidth]{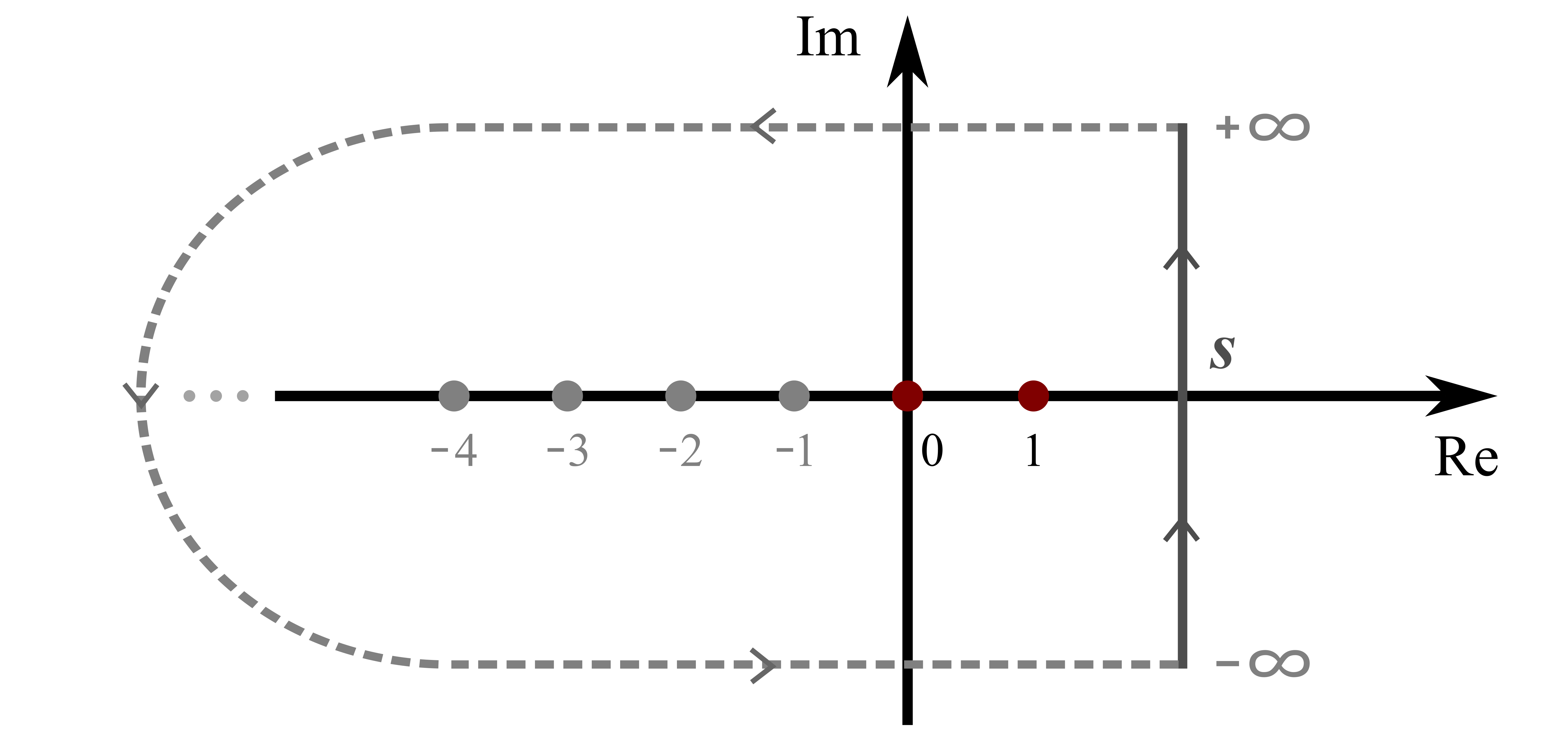}  
  \caption{Contour plot with poles for the U(1) modes specific heat. The $\zeta(s)$ contributes the pole at $s=1$, $\zeta(s+1)$ contributes pole at $s=0$, and $\Gamma(s+2)$ contribute poles at $s=-2,-3,-4...$} 
  \label{fig:CP}
\end{figure}

The chiral bosonic (e.g., the edge modes in the Laughlin states) edge heat capacity reads:
\begin{equation}
\begin{aligned}
    C_{L}=& \sum_{j=1}^{\infty} (j\gamma)^2 \frac{e^{-j\gamma}}{(1-e^{-j\gamma})^2},
\end{aligned}
\label{eqnExactC}
\end{equation}
where $\gamma = \beta \alpha_1$. Here, the ``amplitude" $\lambda_j=1$, the "frequency" $\omega_j=j$ and the function $f(x)=x^2 \frac{e^{-x}}{(1-e^{-x})^2}$. This can thus transform into:
\begin{equation}
    \left(\sum_j j^{-s}\right) \cdot f^{\star}(s) =  \zeta(s) \cdot \zeta(s+1) \cdot \Gamma(s+2) ,
\label{eqn:LaughlinPoles}
\end{equation}
where we have used the properties of the Mellin transform:
\begin{equation}
\begin{aligned}
    f'(x) \stackrel{\mathcal{M}}{\mapsto}& -(s-1) \cdot f^{\star}(s-1)\\
    x^\nu f(x) \stackrel{\mathcal{M}}{\mapsto}& f^{\star}(s+\nu),
\end{aligned}
\end{equation}
and considered the following properties:
\begin{equation}
\begin{aligned}
\Gamma(s+1)
=&\int_0^{\infty} t^s e^{-t} d t =\left.t^s\left(-e^{-t}\right)\right|_0 ^{\infty}-\int_0^{\infty}\left(-e^{-t}\right) s t^{s-1} d t =s \int_0^{\infty} t^{s-1} e^{-t} d t=s \cdot \Gamma(s).
\end{aligned}
\end{equation}
Here, $\zeta(s)$ is the Riemann zeta function (which has a pole when the argument is $1$) and $\Gamma(s)$ is the Gamma function (which has a pole when the argument is $0$ or negative integers). Therefore, Eq.~\ref{eqn:LaughlinPoles} contains poles at $s=1,0,-2,-3...$ (see Fig.~\ref{fig:CP}). However, we only need to consider the poles at $s = 0,1 $, since the Riemann zeta function is zero for negative even integers. By using the residue theorem, we can thus obtain the heat capacity in the asymptotic limit, which gives the same result as the first approach and thus shows the flexibility of the Mellin transform:
\begin{equation}
\begin{aligned}
    C_{L} &= \sum_{n=-1}^{1} \ \text{Res}[\zeta(s) \ \zeta(s+1) \Gamma(s+2) \cdot x^{-s},n]\\
    &= \zeta(-1) \ \zeta(0) \
     \Gamma(1) \cdot x+\zeta(0) \cdot \text{Res}[\zeta(s+1),s=0] \cdot \Gamma(2) + \text{Res}[\zeta(s),s=1]\cdot \zeta(2) \ \Gamma(3)\cdot x^{-1}\\
    &= \frac{x}{24} -\frac{1}{2} + \frac{\pi^2}{3x} \sim \frac{\pi^2}{3x}\bigg(1-\frac{3}{2\pi^2}x + \mathcal{O}(x^2)\bigg) ,
\end{aligned}
\end{equation}
where we only keep the lowest-ordered correction term.

\subsection{Majorana fermionic edge mode}

\begin{table}[t]
  \centering
  \renewcommand{\arraystretch}{1.8}
  \setlength{\tabcolsep}{6pt}
  \resizebox{\textwidth}{!}{
  \begin{tabular}{c|cc|ccc|cc}
    \hline\hline
     &
      \multicolumn{2}{c|}{\textbf{Thin Torus}} &
      \multicolumn{3}{c|}{\textbf{Bulk TQFT/Edge CFT}} &
      \multicolumn{2}{c}{\textbf{Disk}} \\ \hline
    \textbf{Boundary} &
      \multicolumn{1}{c|}{Occupations} &
      Wave functions &
      \multicolumn{1}{c|}{Primary Fields} &
      \multicolumn{1}{c|}{Parity} &
      Characters $(r=1,2)$ &
      \multicolumn{1}{c|}{$N \in \mathbb{N}$} &
      $N \rightarrow \infty$ \\ \hline
    \multirow{4}{*}{\textbf{NS sector}} &
      \multicolumn{1}{c|}{11001100 $\cdots$} &
      \multirow{4}{*}{\begin{tabular}[c]{@{}c@{}}  $\operatorname{Pf}\left(\frac{\vartheta_a\left(z_i-z_j\right)}{\vartheta_1\left(z_i-z_j\right)}\right) \Psi_{1 / 2}^{(t)}$, \\ $a \in \{3,4\}$ \\ \end{tabular}} &
      \multicolumn{1}{c|}{$e^{i \phi / \sqrt{2}}$} &
      \multicolumn{1}{c|}{\multirow{2}{*}{P=1}} &
      $\chi_0^{\text {Ising }} \cdot \left(\chi_{r / 2}^{+} +\chi_{r / 2}^{-} \right)$ &
      \multicolumn{1}{c|}{\multirow{2}{*}{$Z_{MR}^{\mathbf{1}}$}} &\multirow{6}{*}{$Z_{\text{MR}}^{(\infty)}$}
      \\[-0.2em]
     &
      \multicolumn{1}{c|}{00110011 $\cdots$} &
       & 
      \multicolumn{1}{c|}{$e^{-i \phi / \sqrt{2}}$} &
      \multicolumn{1}{c|}{} & $\chi_{1/2}^{\text {Ising }} \cdot \left(\chi_{r / 2}^{+} -\chi_{r / 2}^{-} \right)$
       &
      \multicolumn{1}{c|}{} &
      \\ \cline{5-7}
     &
      \multicolumn{1}{c|}{10011001 $\cdots$} &
       &
      \multicolumn{1}{c|}{$\psi \cdot e^{i \phi / \sqrt{2}}$} &
      \multicolumn{1}{c|}{\multirow{2}{*}{P=-1}} &
      $\chi_{1/2}^{\text {Ising }} \cdot \left(\chi_{r / 2}^{+} +\chi_{r / 2}^{-} \right)$  & 
      \multicolumn{1}{c|}{\multirow{2}{*}{$Z_{MR}^\psi$}} &
      \\[-0.2em]
     &
      \multicolumn{1}{c|}{01100110 $\cdots$} &
       &
      \multicolumn{1}{c|}{$\psi \cdot e^{-i \phi / \sqrt{2}}$} &
      \multicolumn{1}{c|}{} & $\chi_{0}^{\text {Ising }} \cdot \left(\chi_{r / 2}^{+} - \chi_{r / 2}^{-} \right)$
       &
      \multicolumn{1}{c|}{} &
      \\ \cline{1-7} 
    \multirow{2}{*}{\textbf{R sector}} &
      \multicolumn{1}{c|}{10101010 $\cdots$} &
      \multirow{2}{*}{\begin{tabular}[c]{@{}c@{}}$\operatorname{Pf}\left(\frac{\vartheta_2\left(z_i-z_j\right)}{\vartheta_1\left(z_i-z_j\right)}\right) \Psi_{1 / 2}^{(t)}$\\ \end{tabular}} &
      \multicolumn{1}{c|}{\multirow{2}{*}{\begin{tabular}[c]{@{}c@{}}$\sigma \cdot e^{i \phi /(2 \sqrt{2})}$\\ $\sigma \cdot e^{-i \phi /(2 \sqrt{2})}$\end{tabular}}} &
      \multicolumn{1}{c|}{\multirow{2}{*}{--}} & \multirow{2}{*}{\begin{tabular}[c]{@{}c@{}} $\chi_{1 / 16}^{\text {Ising }} \cdot \chi_{(r+1 / 2) / 2}^{+}$ \\ \end{tabular}} 
       &
      \multicolumn{1}{c|}{\multirow{2}{*}{$Z_{MR}^\sigma$}} &
      \\[-0.2em]
     &
      \multicolumn{1}{c|}{01010101 $\cdots$} &  &
      \multicolumn{1}{c|}{} &
      \multicolumn{1}{c|}{} &
       &
      \multicolumn{1}{c|}{} &
      \\ \hline\hline
  \end{tabular}%
  }
  \caption{\textbf{Correspondence between thin-torus occupation patterns, conformal field theory (CFT) descriptions of the Moore-Read (MR) Pfaffian edge, and disk partition functions}. 
  In the thin-torus (Tao-Thouless) limit, the six topologically distinct MR ground states appear as crystalline occupation patterns such as $1100\,1100\cdots$ or $1010\,1010\cdots$, where the quasihole and quasiparticle excitations can be interpreted as domain walls between these different patterns (or ``vacua'')~\cite{bergholtz06pfaffian,Ardonne2008}. 
  Their real-space wave functions are Pfaffians of Jacobi theta functions $\vartheta_a(z)$ multiplied by the bosonic Laughlin state at $\nu=1/2$ on torus, $\Psi_{1/2}^{(t)}=\prod_{i<j} \vartheta_1\left(\left.\frac{z_i-z_j}{L_1} \right\rvert\, i \frac{L_2}{L_1}\right)^2$, where $L_1, L_2$ are the periods of torus, and the Pfaffian factor encodes pairing correlations~\cite{ milovanovic1996edge,read1999}. 
  In the edge CFT description, the MR state factorizes into a neutral Majorana fermion (Ising CFT with primary fields $\mathbf{1}$, $\psi$, and $\sigma$) and a charged $U(1)$ boson $e^{\pm i\phi/\sqrt{2}}$~\cite{milovanovic1996edge, sohal2020entanglement}. 
  The Neveu-Schwarz (NS) sector accommodates the vacuum $\mathbf{1}$ and fermion $\psi$ fields with even/odd fermion parity, while the Ramond (R) sector hosts the spin field $\sigma$ combined with half-charge bosonic operators $e^{\pm i\phi/2\sqrt{2}}$. 
  The disk partition functions $Z_{MR}^{\mathbf{1}}$, $Z_{MR}^{\psi}$, and $Z_{MR}^{\sigma}$ in the main text arise as characters of the corresponding sectors. They are built from one chiral edge with (i) Ising characters $\chi_{h}^{\mathrm{Ising}}(\tau)$ ($2 \pi \tau = i \beta$), projected to a fixed particle number $N$, which generate the Virasoro towers of the primary fields with conformal weights $h=0,\,1/2,\,1/16$, and (ii) $U(1)_2$ characters $\chi^{+}_{r/2}(\tau) \pm \chi^{-}_{r/2}(\tau)$, which describe charge sectors distinguished by the fermion-parity of the edge excitations~\cite{ ino1999multiple,Dong2008} but gets trivialized by the particle number projection on a disk. In the thermodynamic limit $N\to\infty$, all sectors yield the same universal thermal Hall conductance $\kappa=c \,\kappa_0 T$ with central charge $c=3/2$, but they differ by finite-size corrections controlled by the conformal weights of the corresponding primaries~\cite{kane1997quantized, cappelli2002thermal}. It is therefore necessary to carefully account for sector dependence when analyzing finite droplets of the Moore–Read state. An illustration of the mapping between cylinder and disk geometry is shown in Fig.~\ref{fig:cylinder_to_disk}}
  \label{tab:table_dict}
\end{table}

In this subsection, we first show that the partition function $Z_{MR}^{(\infty)}$ in the main text can be interpreted as a linear combination of $Z_{MR}^{\mathbf{1}}$ and $Z_{MR}^{\psi}$ within the language of integer partitions. We then derive the asymptotic limit of the thermal Hall conductance as $\beta\alpha_1 \rightarrow 0$ for different sectors of the Majorana fermion edge mode. Finally, Table~\ref{tab:table_dict} provides a detailed correspondence between the partition functions employed in this work, the microscopic wave functions, and the conformal field theory (CFT) characters.

The partition function $Z^{(\infty)}_{MR}$ can be written as:
\begin{equation}
\begin{aligned}
    Z_{MR}^{(\infty)} =& \prod_{j=0}^{\infty} \left(1+q^{j+1/2} \right)=\underbrace{(1+q^2+q^3+2q^4+...)}_{\text{odd number of distinct odd parts}}+\underbrace{(q^{1/2} +q^{3/2}+q^{5/2}+...)}_{\text{even number of distinct odd parts}}\\
    =& \frac{1}{2}\left[\prod_{j=0}^{\infty} \left(1+q^{j+1/2} \right)+\prod_{j=0}^{\infty} \left(1+q^{j+1/2} \right) \right] + \frac{1}{2}\left[\prod_{j=0}^{\infty} \left(1-q^{j+1/2} \right)-\prod_{j=0}^{\infty} \left(1-q^{j+1/2} \right) \right]\\
    =& \frac{1}{2}\left[\prod_{j=0}^{\infty} \left(1+q^{j+1/2} \right)+\prod_{j=0}^{\infty} \left(1-q^{j+1/2} \right) \right] + \frac{1}{2}\left[\prod_{j=0}^{\infty} \left(1+q^{j+1/2} \right)-\prod_{j=0}^{\infty} \left(1-q^{j+1/2} \right) \right]\\
    =& Z_{MF}^{\mathbf{1}} + Z_{MF}^{\psi}
\end{aligned}
\end{equation}
where one can assume that the series is absolutely convergent, so that the infinite product can be rearranged freely. In this case, the partition function $Z_{MR}^{(\infty)}$ can be decomposed into two distinct generating functions: (1) those corresponding to distinct odd parts with an odd number of parts, and (2) those corresponding to distinct odd parts with an even number of parts. These generating functions correspond respectively to $Z_{MF}^{\mathbf{1}}$ and $Z_{MF}^{\mathbf{\psi}}$. Hence, $Z_{MR}^{(\infty)}$ serves as the generating function for distinct odd parts. For the $\sigma$-sector partition function,  
$Z_{MF}^{\sigma} = \prod_{j=0}^{\infty} \left(1+q^j\right)$, 
which represents the generating function for partitions into distinct parts (equivalently, partitions into odd parts).

\begin{figure}
  \includegraphics[width=0.6\linewidth]{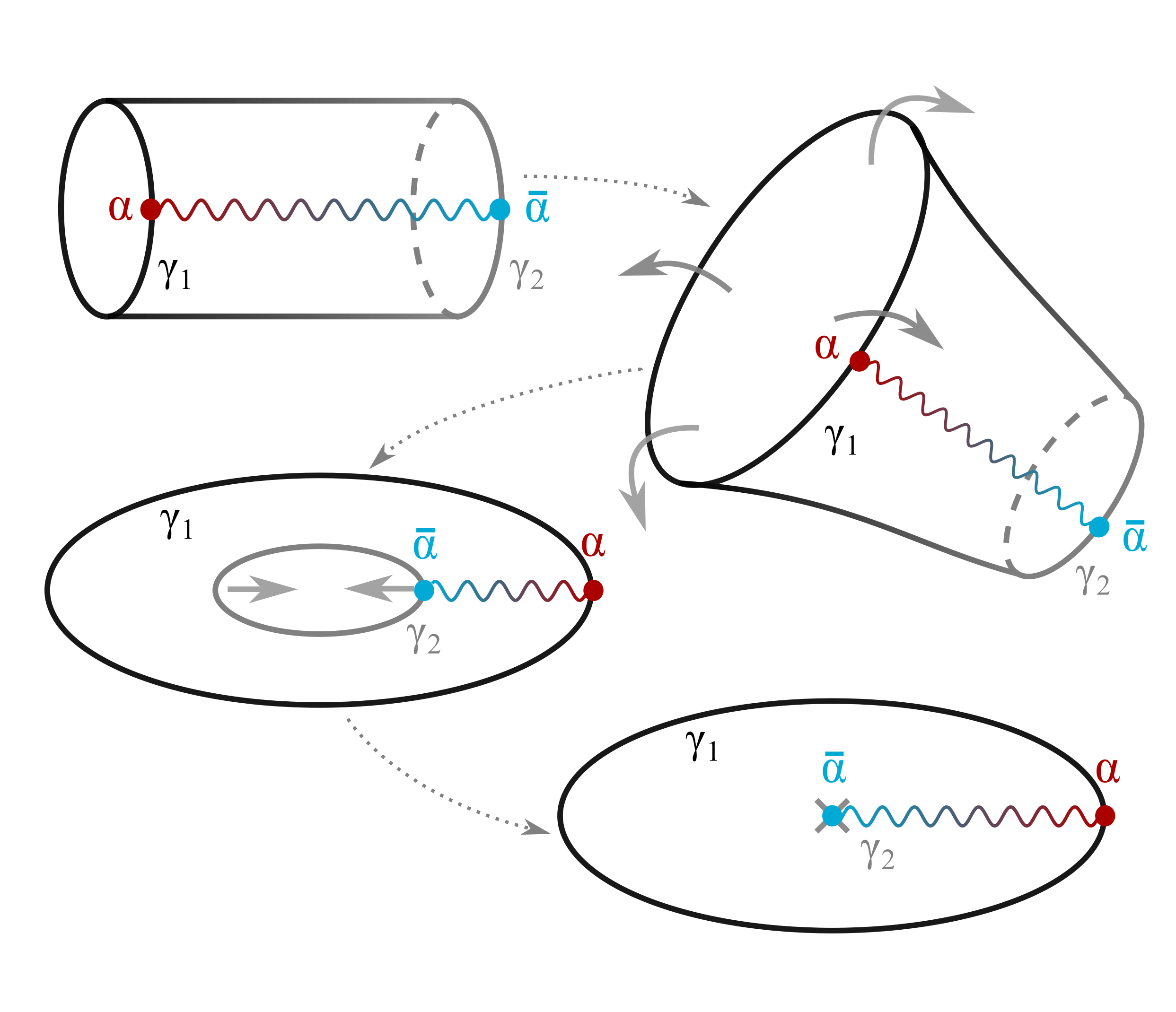}  
  \caption{\textbf{Mapping from cylinder to disk geometry}. On a cylinder with two chiral edges $\gamma_{1,2}$, inserting a flux through the hole is equivalent to nucleating a conjugate anyon pair $(a,\bar a)$ in the bulk, and one can drag them to opposite boundaries, represented by an open Wilson line stretching between $\gamma_1$ and $\gamma_2$. Here, the bulk-edge correspondence appears as a \textit{gluing condition} that originates from electron locality and enforces conjugate anyon charges on the two edges. Shrinking $\gamma_2$ to a point in the bulk maps the cylinder to a disk, where the Wilson line now terminates at the single boundary $\gamma_1$. In this limit, the cylinder gluing condition gives the bulk-edge correspondence on the disk, i.e., the bulk fixes the topological sector of the remaining edge, thereby determining the Majorana boundary condition (NS or R), the parity, and the $U(1)_2$ charge sector of the Moore-Read edge CFT. The resulting disk partition functions are the corresponding character combinations \cite{sohal2020entanglement}.}
  \label{fig:cylinder_to_disk}
\end{figure}

Let us first derive the asymptotic limit of the heat capacity for the $Z^{(\infty)}_{MF}$. Given this partition function, we can derive the exact form of the heat capacity as:
\begin{equation}
\begin{aligned}
    C_{MF}^{\infty}= \sum_{n=0}^{\infty} \left[\gamma \left(n+\frac{1}{2}\right)\right]^2 \frac{e^{-\gamma\left(n+\frac{1}{2}\right)}}{[1+e^{-\gamma(n+\frac{1}{2})}]^2}
\end{aligned}
\end{equation}
where $\gamma=\beta \alpha_1 $. Here, the ``amplitude'' $\lambda_j=1$, the ``frequency'' $\omega_j=j+\frac{1}{2}$ and the function $f(x)=x^2 \frac{e^{-x}}{(1+e^{-x})^2}$. This can thus transform into:
\begin{equation}
\begin{split}
    & \left(\sum_j (j+\frac{1}{2})^{-s}\right) \cdot f^{\star}(s) = (s+1)\cdot(2^s+2^{-s}-2)\cdot \Gamma(s)\cdot \zeta(s) \cdot \zeta(s+1)
\end{split}
\end{equation}
By using the residue theorem, the heat capacity of the Majorana Fermion mode in asymptotically limit ($\beta \alpha_1 \rightarrow 0$) is thus:
\begin{equation}
    C_{MF}^{(\infty)} = \frac{\pi^2}{3x}\left(\frac{1}{2} + \mathcal{O}(x^3) \right) 
\end{equation}
in which we can see that the leading order correction term vanishes (compare with the one in chiral boson). 

Next, we derive the asymptotic limit of the heat capacity for the  $\mathbf{1}$-sector of the Majorana fermion edge mode. Notice that the partition function $Z^{\mathbf{1}}_{MF}$ can be written as:
\begin{equation}
    Z_{MF}^{\mathbf{1}} = \frac{1}{2} \left[\prod_{j=0}^{\infty} \left(1+q^{j+1/2} \right) +\prod_{j=0}^{\infty} \left(1-q^{j+1/2} \right) \right] = \prod_{n=1}^{\infty}\frac{\left(1+q^{8n-3} \right)\left(1+q^{8n-5} \right)\left(1-q^{8n} \right)}{\left(1-q^{2n} \right)},
\label{eqn:MFIdentity}
\end{equation}
where the second equality of Eq.~\ref{eqn:MFIdentity} can be found in Ref.\cite{slater1952further}. Let us first focus on the term with $\prod_{n=1}^{\infty}\left(1+q^{8n-3} \right)$. The heat capacity of this term reads:
\begin{equation}
    C_{8n-3} = \sum_{n=1}^{\infty} \left[\gamma(8n-3) \right]^2 \frac{e^{-\gamma (8n-3)}}{\left[1-e^{-\gamma (8n-3)} \right]^2},
\end{equation}
in which we recognize the amplitude $\lambda_j=1$, $\omega_j=8j-3$, and the function $f(x)=x^2 e^{-x}/(1-e^{-x})^2$. Hence, by using the Mellin transform, we obtain:
\begin{equation}
    \left[\sum_{j=1}^{\infty} (8j-3)^{-s} \right]\cdot f^{\star}(s) = 8^{-s} \cdot \zeta\left(s,\frac{5}{8}\right)\cdot (s+1)\cdot \Gamma(s+1) \cdot \zeta(s+1)\cdot (1-2^{-s}).
\end{equation}
Based on the residue theorem, the heat capacity contributed by this term reads:
\begin{equation}
    C_{8n-3} = \sum_{n=-1}^{1} \text{Res} \left[ 8^{-s} \cdot \zeta\left(s,\frac{5}{8}\right)\cdot (s+1)\cdot \Gamma(s+1) \cdot \zeta(s+1)\cdot (1-2^{-s})\cdot x^{-s},n\right]= \frac{\pi^2}{48x} + \mathcal{O}(x^2).
\end{equation}
A similar approach can be done for the next three terms, in which we find that the heat capacity contributed by these terms (after Mellin transformed) reads:
\begin{equation}
    C_{8n-5} = \frac{\pi^2}{48 x} +\mathcal{O}(x^2),\quad
    C_{8n}   = -\frac{\pi^2}{24x} +\frac{1}{2}+\mathcal{O}(x^2),\quad
    C_{2n}   = \frac{\pi^2}{6x} -\frac{1}{2}+\mathcal{O}(x^2).
\end{equation}
Hence, by adding up all the heat capacity contributed by these four terms, we got the asymptotic limit of the heat capacity for the Majorana Fermion edge state:
\begin{align}
    C^{\mathbf{1}}_{MF} &= \frac{\pi^2}{6x} + \mathcal{O}(x^2) = \frac{\pi^2}{3x} \left(\frac{1}{2} +\mathcal{O}(x^3)\right),
\end{align}
which gives us the correct central charge $c=1/2$, and we can see that there is no linear order correction term in $x$, and hence $\beta \alpha_1$. 

Finally, we derive the asymptotic limit of the heat capacity for $\sigma$ sector. The partition function reads: $Z^\sigma_{MF}=\prod_{j=0}^{\infty}(1+q^j)$, in which we can derive the heat capacity to be:
\begin{equation}
\begin{aligned}
    C_{MF}^{\sigma}= \sum_{j=1}^{\infty} \left(\gamma j\right)^2 \frac{e^{-\gamma j}}{(1+e^{-\gamma j})^2}
\end{aligned}
\end{equation}
where $\gamma=\beta \alpha_1 $. Here, the ``amplitude'' $\lambda_j=1$, the ``frequency'' $\omega_j=j$ and the function $f(x)=x^2 \frac{e^{-x}}{(1+e^{-x})^2}$. This can thus transform into:
\begin{equation}
\begin{split}
    & \left(\sum_j j^{-s}\right) \cdot f^{\star}(s) = \zeta(s) \cdot(s+1)(1-2^{-s})\Gamma(s+1)\zeta(s+1)
\end{split}
\end{equation}
By using the residue theorem, the heat capacity of the Majorana Fermion mode in asymptotically limit ($\beta \alpha_1 \rightarrow 0$) is thus:
\begin{equation}
    C_{MF}^{\sigma} = \frac{\pi^2}{3x}\left(\frac{1}{2} + \mathcal{O}(x^3) \right) 
\end{equation}
in which we can again see that the leading order correction term vanishes. From all the calculation above, we can see that regardless of the sector, the central charge for the Majorana fermion mode is $c= 1/2$, as predicted by CFT.

\subsection{Gaffnian edge mode}
The partition function for the edge of the Gaffnian state \cite{read2009conformal} reads:
\begin{align}
    Z_G &= \frac{1}{(q)_{\infty}} \sum_{n=0}^{\infty}\frac{q^{n(n+1)}}{(q)_{2\infty}} = \frac{1}{(q)_{\infty}} \prod_{n=1}^{\infty}\frac{1}{\left[1-(-q)^{5n-4}\right]\left[1-(-q)^{5n-1}\right](1-q^{2n-1})},
\label{eqn:GaffnianPartitionFunction}
\end{align}
where the second equality in Eq.~\ref{eqn:GaffnianPartitionFunction} can be found in Ref.\cite{baxter2016exactly}. The $1/(q)_{\infty}$ term is the Abelian chiral bosonic mode and will give the central charge of $c=1$. The remaining factor describes the minimal model of M(5,3), which is non-Abelian. Such a minimal model has the central charge of $c=0.6$, which is predicted from CFT. Here, we will only focus on the non-Abelian part of the Eq.~\ref{eqn:GaffnianPartitionFunction}. 

Let us first focus on the term with $\prod_{n=1}^{\infty}(1-q^{2n-1})^{-1}$. The heat capacity of this term reads:
\begin{equation}
    C_{2n-1} = \sum_{n=1}^{\infty} \left[\gamma(2n-1)\right]^2 \frac{e^{-\gamma(2n-1)}}{\left[ 1-e^{-\gamma(2n-1)}\right]^2},
\end{equation}
in which we recognize the amplitude $\lambda_j=1$, $\omega_j=2j-1$, and the function $f(x)=x^2e^{-x}/(1-e^{-x})^2$. Hence, by using the Mellin transform, we obtain:
\begin{equation}
    \left[\sum_j (2j-1)^{-s}\right] \cdot f^{\star}(s) 
    = 2^{-s} \cdot (2^s-1) \cdot \zeta(s) \cdot(s+1)\cdot \Gamma(s+1) \cdot \zeta(s+1),
\end{equation}
By using the residue theorem, the heat capacity contributed by this term reads:
\begin{align}
    C_{2n-1} &= \sum_{n=-1}^{1} \text{Res} \left[ 2^{-s} \cdot (2^s-1) \cdot \zeta(s) \cdot(s+1)\cdot \Gamma(s+1) \cdot \zeta(s+1)\cdot x^{-s},n\right]= \frac{\pi^2}{6x}.
\end{align}

Now we consider the term with $\prod_{n=1}^{\infty}\left[1-(-q)^{5n-4}\right]^{-1}$, which contribute heat capcity of:
\begin{align}
    C_{5n-4} &= \sum_{n=1}^{\infty} \left[\gamma(5n-4)\right]^2 \frac{(-e)^{-\gamma(5n-4)}}{\left[ 1-(-e)^{-\gamma(5n-4)}\right]^2}\\
    &= -\sum_{n\in odd}^{\infty} \left[\gamma(5n-4)\right]^2 \frac{e^{-\gamma(5n-4)}}{\left[ 1+e^{-\gamma(5n-4)}\right]^2}+\sum_{n\in even}^{\infty} \left[\gamma(5n-4)\right]^2 \frac{e^{-\gamma(5n-4)}}{\left[ 1-e^{-\gamma(5n-4)}\right]^2}. \label{eqn:5n-4}
\end{align}
The first term in Eq.~\ref{eqn:5n-4} can be Mellin transformed into:
\begin{equation}
    \left[\sum_{j\in odd} (5j-4)^{-s}\right] \cdot f^{\star}(s) 
    = 10^{-s} \cdot \zeta\left(s,\frac{1}{10}\right)  \cdot(s+1)\cdot \Gamma(s+1) \cdot \zeta(s+1)\cdot (1-2^{-s}),
\label{eqn:5j-4odd}
\end{equation}
whereas the second term in Eq.~\ref{eqn:5n-4} can be Mellin transformed into:
\begin{equation}
    \left[\sum_{j\in even} (5j-4)^{-s}\right] \cdot f^{\star}(s) 
    = 10^{-s} \cdot \zeta\left(s,\frac{3}{5}\right)  \cdot(s+1)\cdot \Gamma(s+1) \cdot \zeta(s+1).
\label{eqn:5j-4even}
\end{equation}
By doing the residue theorem of both the contributions from both Eq.~\ref{eqn:5j-4odd} and Eq.~\ref{eqn:5j-4even}, the heat capacity contributed is thus:
\begin{align}
    C_{5n-4} &= -\frac{\pi^2}{60 x}- \mathcal{O}(x^2) + \frac{\pi^2}{30x} -\frac{1}{10} -\mathcal{O}(x^2)= \frac{\pi^2}{60x} -\frac{1}{10} - \mathcal{O}(x^2).
\end{align}

Now, we consider the last term with $\prod_{n-1}^{\infty}\left[1-(-q)^{5n-1}\right]^{-1}$, which contribute heat capacity of:
\begin{align}
    C_{5n-1} &= \sum_{n=1}^{\infty} \left[\gamma(5n-1)\right]^2 \frac{(-e)^{-\gamma(5n-1)}}{\left[ 1-(-e)^{-\gamma(5n-1)}\right]^2}\\
    &= \sum_{n\in odd}^{\infty} \left[\gamma(5n-1)\right]^2 \frac{e^{-\gamma(5n-1)}}{\left[ 1-e^{-\gamma(5n-1)}\right]^2}-\sum_{n\in even}^{\infty} \left[\gamma(5n-1)\right]^2 \frac{e^{-\gamma(5n-1)}}{\left[ 1+e^{-\gamma(5n-1)}\right]^2}. \label{eqn:5n-1}
\end{align}
The first term in Eq.~\ref{eqn:5n-1} can be Mellin transformed into:
\begin{equation}
    \left[\sum_{j\in odd} (5j-1)^{-s}\right] \cdot f^{\star}(s) 
    = 10^{-s} \cdot \zeta\left(s,\frac{2}{5}\right)  \cdot(s+1)\cdot \Gamma(s+1) \cdot \zeta(s+1)\cdot 
\label{eqn:5j-1odd}
\end{equation} 
whereas the second term in Eq.~\ref{eqn:5n-1} can be Mellin transformed into:
\begin{equation}
    \left[\sum_{j\in even} (5j-4)^{-s}\right] \cdot f^{\star}(s) 
    = 10^{-s} \cdot \zeta\left(s,\frac{9}{10}\right)  \cdot(s+1)\cdot \Gamma(s+1) \cdot \zeta(s+1)\cdot(1-2^{-s}).
\label{eqn:5j-1even}
\end{equation}
Finally, by doing the residue theorem, the heat capacity contributed by this term is thus: 
\begin{align}
    C_{5n-1} &= \frac{\pi^2}{30 x}+\frac{1}{10}+\mathcal{O}(x^2) - \frac{\pi^2}{60x}  -\mathcal{O}(x^2)= \frac{\pi^2}{60x} +\frac{1}{10} - \mathcal{O}(x^2).
\end{align}
Hence, by adding up all the heat capacity contributed by these three terms, we got the asymptotic limit of the heat capacity for the non-Abelian part of the Gaffnian state:
\begin{align}
    C_{NA} = \frac{\pi^2}{5x} -\mathcal{O}(x^2) = \frac{\pi^2}{3x} \left(\frac{3}{5}-\mathcal{O}(x^3)\right),
\end{align}
which gives us the correct central charge $c=0.6$, and there is no linear order correction term in $\beta \alpha_1$, just like the case in the Majorana Fermion, again indicating the robustness against temperature for the non-Abelian modes.

\subsection{Chiral bosonic edge mode with non-zero self-energy \label{SectionLaughlinNonZero}}
We first show that Eq.~\ref{eqn:Laughlin self energy} is the correct partition function that could capture the effect of non-zero self-energy for the chiral bosonic mode. Recall that we have assumed that the energy cost of each flux insertion has a constant energy $\mu$, and we denote $q\equiv e^{-\beta\alpha_1}$ and $t=e^{-\beta \mu}$. We start by expanding the RHS of Eq.~\ref{eqn:Laughlin self energy}:
\begin{equation}
\begin{aligned}
    \prod_{n=1}^{\infty} \frac{1}{1-tq^n} &= 1+ \underbrace{tq}_{\Delta m=1} + \underbrace{t^2 q^2+tq^2}_{\Delta m =2} + \underbrace{t^3q^3 + t^2q^3 + tq^3}_{\Delta m=3} 
    + \underbrace{t^4q^4 + t^3q^4 + t^2q^4 +tq^4+t^2q^4}_{\Delta m =4} + \cdots.
\end{aligned}
\end{equation}
Let us take the three terms in the $\Delta m =3$ sector as an example, the $t^3 q^3$ term corresponds to the quasihole state with three quasiholes formation (hence the energy cost is $3\mu$); the $t^2q^3$ term corresponds to the quasihole state with two quasiholes formation and the $tq^3$ term corresponds to the quasihole state with one quasihole formation. One can check that the remaining terms for the other $\Delta m$ sector are also compatible with the one in the LHS of Eq.~\ref{eqn:Laughlin self energy}. Notice that if $\mu =0$ (i.e $t=1$), which is the ideal case we have considered in the previous section, we recover the $p(\Delta m)$ degeneracy.

From this partition function, we can easily compute the specific heat as:
\begin{equation}
    C_{L,qh} = k_B \sum_{j=1}^{\infty} \left[\gamma \left(j+\frac{\mu}{\alpha_1}\right) \right]^2 \frac{e^{\gamma\left(j+\frac{\mu}{\alpha_1}\right)}}{\left[1-e^{\left(j+\frac{\mu}{\alpha_1}\right)} \right]^2},
\end{equation}
where $\gamma = \beta \alpha_1$. We can again recognize that the ``amplitude'' $\lambda_j=1$, the ``frequency'' $\omega_j = j+\frac{\mu}{\alpha_1}$ and the function $f(x) = x^2 \frac{e^x}{(1-e^x)^2}$. This can thus transform into:
\begin{equation}
\begin{split}
    & \left[\sum_j \left(j+\frac{\mu}{\alpha_1}\right)^{-s}\right] \cdot f^{\star}(s) 
    = \zeta\left(s,1+\frac{\mu}{\alpha_1}\right) \cdot(s+1)\cdot\Gamma(x+1)\cdot \zeta(s+1),
\end{split}
\end{equation}
where $\zeta(s,a)$ is the Hurwitz Zeta function with the following important property: $\zeta(0,a)=\frac{1}{2}-a$. By taking the residues of all the poles here, we can obtain the specific heat in the asymptotic limit as:
\begin{equation}
\begin{aligned}
    C_{L,qh} &= \sum_{n} \text{Res} \left[ \zeta\left(s,1+\frac{\mu}{\alpha_1}\right) \cdot(s+1)\cdot\Gamma(x+1)\cdot \zeta(s+1)\cdot x^{-s},n\right]\\
    &= \frac{\pi^2}{3x}-\frac{1}{2}-\frac{\mu}{\alpha_1} + \mathcal{O}(x,\mu^2)=\frac{\pi^2}{3\beta \alpha_1}\left[1-\frac{3}{2\pi^2}\beta (\alpha_1 + 2\mu)+\mathcal{O}(\beta^2,\alpha_1^2,\mu^2) \right],
\end{aligned}
\end{equation}
in which we obtain the correction terms that are linear to $\beta \alpha_1$ and $\beta \mu$.

\subsection{Majorana Fermion edge mode with non-zero self-energy}
By considering the non-zero self-energy, the partition function for the Majorana fermion edge mode reads:
\begin{equation}
    Z_{MF,qh} = \prod_{n = 0}^{\infty}(1+q^{n+1/2}t^{1/2})
\end{equation}
in which we can easily derive the specific heat to be:
\begin{equation}
    C_{MF,qh} = k_B \sum_{j=0}^{\infty} \left[\gamma\left(j+\frac{1}{2}\right)+\frac{\beta \mu}{2} \right]\frac{e^{\gamma(j+\frac{1}{2})}e^{\frac{\beta\mu}{2}}}{\left(1 + e^{\gamma(j+\frac{1}{2})}e^{\frac{\beta \mu}{2}} \right)^2}
\end{equation}
where $\gamma=\beta \alpha_1$. We recognize that the ``amplitude'' $\lambda_j=1$, the ``frequency'' $\omega_j=\left(j+\frac{1}{2}\right)+\frac{\mu}{2\alpha_1}$ and the function $f(x)=x^2 \frac{e^x}{(1+e^x)^2}$. Hence, we can perform the Mellin transform:
\begin{equation}
    \left[\sum_{j}\left(j+\frac{1}{2}+\frac{\mu}{2\alpha_1} \right)^{-s} \right]\cdot f^{\star}(s) =\zeta\left(s,\frac{1}{2}+\frac{\mu}{2\alpha_1} \right) (s+1)(1-2^{-s})\Gamma(s+1)\zeta(s+1).
\end{equation}
Note that the Hurwitz Zeta function is related to the Bernoulli number: $\zeta(-s,a)=-B_{s+1}(a)/(s+1)$. Hence, by taking the residues of all the poles here, we obtain the specfic heat in the asymptotic limit ($\beta \alpha_1 \rightarrow 0$) to be:
\begin{equation}
\begin{aligned}
    C_{MF,qh} &= \sum_{n}\text{Res}\left[\zeta\left(s,\frac{1}{2}+\frac{\mu}{2\alpha_1} \right) (s+1)(1-2^{-s})\Gamma(s+1)\zeta(s+1)\cdot x^{-s},n \right]\\
    &= \frac{\pi^2}{6x} + \mathcal{O}(x^2,\mu^2) = \frac{\pi^2}{3\beta \alpha_1}\left(\frac{1}{2} + \mathcal{O}(\beta^3,\alpha_1^3,\mu^2) \right)
\end{aligned}
\end{equation}
where we can see that the correction terms that are linear to $\beta \alpha_1$ and $\beta \mu$ vanish, once again showing that there is an intrinsic robustness for Majorana fermion edge mode

\section{Equivalence between microscopic picture and Luttinger liquid formalism \label{AppendixB}}

In this section, we will show that if finite-size effect is taken into account, the THC is no longer a universal quantity under the general dispersion relation. This argument can be reflected if we approximate the discrete summation over $\Delta m$ into a continuous integral. To do this, we expand the exact form of specific heat of the chiral $U(1)$ boson mode by using \textit{Euler-Maclaurin formula}:
\begin{align}
    C_{U(1)} &= \sum_{j=1}^{\infty} \underbrace{(j \beta \alpha_1)^2 \frac{e^{-j\beta \alpha_1}}{(1-e^{-j\beta \alpha_1})^2}}_{f(j)}\\
    &\approx \int_{1}^{\infty} dj f(j) +\frac{f(\infty)-f(1)}{2} + \frac{1}{12}[f'(\infty)-f'(1)] -\frac{1}{720}[f'''(\infty)-f'''(1)] + \cdots. \label{eqn: EulerMaclaurinLaughlin}
\end{align}
To avoid the divergence problem, we will divide an extra factor $C_0=\pi^2/(3\gamma)$ into Eq.~\ref{eqn: EulerMaclaurinLaughlin}. Let us focus just on the first term of Eq.~\ref{eqn: EulerMaclaurinLaughlin}:
\begin{align}
    \frac{C_{U(1)}}{C_0} &\approx \frac{3\gamma}{\pi^2} \int_{1}^{\infty} dj\:\frac{(j\gamma)^2 e^{-j\gamma}}{\left(1-e^{-j\gamma} \right)^2} =\frac{3}{\pi^2} \int_{\gamma}^{\infty} dx \: \frac{x^2 e^{-x}}{\left(1-e^{-x} \right)^2},
\label{eqn:AfterEulerMa}
\end{align}
where we have performed the variable change $x=j\gamma$ in the second equality on Eq.~\ref{eqn:AfterEulerMa}. By comparing Eq.~\ref{eqn:AfterEulerMa} with the continuous case, one can see that Eq.~\ref{eqn:LinearTHC} is its limit at $\gamma \rightarrow 0$: Only in this case, the integrand is $\pi^2/3$ and the universal $c=1$ is recovered. If the second term of Eq.~\ref{eqn:AfterEulerMa} is included, one can show that the first-order correction terms of the Euler-Maclaurin formula at the limit of at $\gamma \rightarrow 0$ $f(\infty) = 0$ and $f(1) \approx \frac{3}{\pi^2}\gamma$, which recover the THC with first-order correction term due to the finite-size effects.

\section{THC under a  More General Dispersion Relation \label{AppendixC}}
In this section, we consider the THC of the Laughlin $\nu = 1/3$ state under a more general dispersion relation ($\epsilon=\alpha_1n+\alpha_2n^2$). Assuming the degeneracy at each angular momentum state $p(n)$ is not affected, the partition function reads:
\begin{equation}
    Z = \sum_{n=0}^{\infty} p(n)\cdot e^{-\beta(\alpha_1 n+\alpha_2n^2)},
\end{equation}
where $p(n)$ is the partition number for integer $n$. We can expand the $e^{-\beta \alpha_2 n^2}$ term into linear order term:
\begin{equation}
\begin{aligned}
    Z &= \sum_{n=0}^{\infty} p(n)\cdot e^{-\beta \alpha_1 n} e^{-\beta \alpha_2n^2}
    = \sum_{n=0}^{\infty} p(n)\cdot e^{-\beta \alpha_1 n} (1-\beta \alpha_2n^2)\\
    &=\underbrace{\sum_{n=0}^{\infty} p(n)\cdot e^{-\beta \alpha_1 n}}_{\equiv Z_0} -\beta \alpha_2\underbrace{\sum_{n=0}^{\infty} p(n)\cdot n^2 e^{-\beta \alpha_1 n}}_{\equiv Z_2}
    = Z_0 - \beta\alpha_2 Z_2.
\end{aligned}
\end{equation}
The $\log(Z)$ can be further written as:
\begin{equation}
\begin{aligned}
    \log(Z) &= \log(Z_0) + \log \Big(1-\beta \alpha_2 \frac{Z_2}{Z_0}\Big)\approx \log(Z_0)-\beta \alpha_2 \frac{Z_2}{Z_0}.
\label{eqn:log}
\end{aligned}
\end{equation}
The $\log(Z_0)$ terms can be dealt with by using either the Mellin transform or Euler-Maclaurin expansion. We will focus on the second term in Eq.~\ref{eqn:log}. Typically, we will use \textit{saddle point approximation} to deal with the $\frac{Z_2}{Z_0}$ term. 

By using the Ramanujan partition formula, we can estimate $p(n)\sim \frac{B}{n}e^{A\sqrt{n}}$, where $A=\pi\sqrt{2/3}$ and $B=1/(4\sqrt{3})$. By using this approximation, we have to throw away the $n=0$ term in $Z_2$ summation to avoid divergence. The generating function now looks:
\begin{equation}
    Z_2 = B\sum_{n=1}^{\infty}ne^{A\sqrt{n}-\beta\alpha_1n}=B\sum_{n=1}^{\infty}ne^{\Phi(n)},
\end{equation}
where we have define $\Phi(n)=A\sqrt{n}-\beta\alpha_1n$. We now make use of saddle point approximation, that is, to find the $n=n^*$ that makes $\Phi'(n)=0$, so that we can estimate:
\begin{equation}
    \Phi(n)=\Phi(n^*)-\frac{1}{2}\Phi''(n^*)(n-n^*)^2 + \mathcal{O} \left((n-n^*)^3 \right).
\end{equation}
After some calculation we find $n^*=\frac{\pi^2}{6\beta^2\alpha_1^2}$ and $\Phi(n^*)=\frac{\pi^2}{6\beta\alpha_1}$. With this, we have:
\begin{equation}
\begin{aligned}
    Z_2 &= B\int dn \: n \: e^{\Phi(n)}
    = Bn^*e^{\Phi(n^*)}\int dn \: e^{-\frac{1}{2}\Phi''(n^*)(n-n^*)^2}.
\label{eqn:z2}
\end{aligned}
\end{equation}
We can do the same procedures for $Z_0$ and obtain:
\begin{equation}
    Z_0 = \frac{B}{n^*}e^{\Phi(n^*)}\int dn \: e^{-\frac{1}{2}\Phi''(n^*)(n-n^*)^2}.
\label{eqn:z0}
\end{equation}
The integrals in Eq.~\ref{eqn:z2} and Eq.~\ref{eqn:z0} are just a Gaussian integral, and both integrals have the same result. Thus, we have:
\begin{equation}
    \frac{Z_2}{Z_0} = \frac{n^*}{(\frac{1}{n^*})} = (n^*)^2 = \frac{\pi^4}{36\beta^4\alpha_1^4}.
\end{equation}
With this, we can go back to the Eq.~\ref{eqn:log} to get the heat capacity $C=\beta^2\frac{\partial^2\log(Z)}{\partial\beta^2}$, and we find:
\begin{equation}
    C = \frac{\pi^2}{3\beta\alpha_1}\bigg(1-\underbrace{\frac{3}{2\pi^2}\beta\alpha_1}_{\text{from }Z_0}-\underbrace{\pi^2\frac{\alpha_2}{\alpha_1^3 \beta^2}}_{\text{from }\frac{Z_2}{Z_0}}\bigg).
\end{equation}
Interestingly, we find that by considering an extra quadratic dispersion, the heat capacity has a cubic dependence on temperature $T$.

\section{Heat capacity with nonlinear dispersion\label{AppendixD}}
In this section, we will show the exact derivations of heat capacity with any power-law dispersions to the leading order. Consider the partition function of Laughlin edge modes under a general dispersion relation ($\epsilon_m = \alpha_k (\Delta m)^k$). The trick to approach this is to use the \textit{Hardy-Ramanujan formula} to write down the asymptotic expression of the unrestricted partition number:
\begin{equation}
p(n) \approx \frac{1}{4 n \sqrt{3}} e^{\pi \sqrt{\frac{2 n}{3}}} \equiv \frac{e^{A \cdot n^{1/2}}}{n},
\end{equation}
where $A \equiv \pi \sqrt{2/3}$. Therefore, the partition function can be written as:
\begin{equation}
Z(\beta)=\sum_{\Delta m=1}^{\infty} p(\Delta m) e^{-\beta (\Delta m)^k} \approx \sum_{\Delta m=1}^{\infty} \frac{e^{A (\Delta m)^{1 / 2}}}{\Delta m} e^{-\beta (\Delta m)^k}.
\end{equation}
Here, we have changed the lower bound of the summation to be $1$ instead of $0$ since the Hardy-Ramanujan approximation is only valid for large $n$, which becomes increasingly accurate as $n$ grows. Therefore, the contribution from $n = 0$ is negligible compared to the contributions from larger $n$ so we took it out from the summation. By approximating the sum as an integral for large $n$, we have: 
\begin{equation}
 Z(\beta) \approx \int_1^{\infty} \frac{e^{A (\Delta m)^{1 / 2}}}{\Delta m} e^{-\beta (\Delta m)^k} d(\Delta m) \equiv \int_1^{\infty}  e^{\Phi(\Delta m)} d(\Delta m).
\end{equation}
And we define the exponent function as:
\begin{equation}
\Phi(\Delta m) \equiv A (\Delta m)^{1 / 2}-\beta (\Delta m)^k-\ln (\Delta m).
\end{equation}
To apply the \textit{saddle-point approximation}, we need to find the value of $\Delta m=\Delta m_0$
where $\Phi(n)$ is maximized:
\begin{equation}
\Phi^{\prime}(\Delta m)=\frac{d \Phi}{d (\Delta m)} \approx \frac{A}{2} (\Delta m)^{-1 / 2}-\beta k (\Delta m)^{k-1},
\end{equation}
where we have neglected the $1/\Delta m$ term, which gives the saddle point as:
\begin{equation}
\Delta m_0=\left(\frac{A}{2 \beta k}\right)^{\frac{1}{k-\frac{1}{2}}}.
\end{equation}
Expanding the exponent function at $\Delta m=\Delta m_0$ we get:
\begin{equation}
\Phi(\Delta m) \approx \Phi\left(\Delta m_0\right)+\frac{1}{2} \Phi^{\prime \prime}\left(\Delta m_0\right)\left(\Delta m-\Delta m_0\right)^2 + \mathcal{O}(\Delta m^3).
\end{equation}
Substituting this into the integral form of the partition function gives:
\begin{equation}
Z(\beta) \approx e^{\Phi\left(\Delta m_0\right)} \int e^{-\frac{1}{2} \Phi^{\prime \prime}\left(\Delta m_0\right)\left(\Delta m-\Delta m_0\right)^2} d(\Delta m).
\end{equation}
If we only keep the leading order term, we have:
\begin{equation}
\ln Z(\beta) \approx \Phi\left(\Delta m_0\right).
\end{equation}
With the approximated form of the logarithmic function of $Z(\beta)$, we have: 
\begin{equation}
E=-\frac{d}{d \beta} \ln Z(\beta) = -\frac{\partial \Phi}{\partial \beta} = \Delta m_0^k =\left(\frac{A}{2 k}\right)^{\frac{k}{k-\frac{1}{2}}} \beta^{-\frac{k}{k-\frac{1}{2}}}.
\end{equation}
Finally, we obtain the specific heat for the Laughlin edge modes at $k$ order dispersion to be:
\begin{equation}
    C_{L} = \frac{2k}{2k-1} \left( \frac{A}{2k} \right)^{\frac{2k}{2k-1}} T^{\frac{1}{2k-1}},
\label{eqn:SaddlePoint}
\end{equation}
which proved our earlier statement. The coefficient of specific heat in Eq. \ref{eqn:SaddlePoint} also fits the numeric results in Fig. \ref{fig:QuadraticHeat}.

We can use the same approach to obtain the specific heat for the Majorana fermion, and it turns out that it has a similar form as the Eq. \ref{eqn:SaddlePoint}, with a difference in $A=\pi \sqrt{1/3}$.

\section{Extract THC from partition functions \label{AppendixE}}
In this section, we showed how to obtain the relation between the THC or thermal current with the partition function $Z$ regardless of the dispersion. We start with the definition of thermal current:
\begin{equation}
J_Q = \sum_{\Delta m=0}^{\infty} v_{m} \frac{\epsilon_{m}}{L}\frac{p(\Delta m) e^{-\beta \epsilon_{m}}}{Z},
\label{eqn:GeneralThermalCurrent(Appendix)}
\end{equation}
where the velocity $v_{m}$ reads:
\begin{equation}
    v_{m} = \frac{\partial \epsilon_{m}}{\partial (\Delta m)} \cdot \frac{2\pi}{L}.
\end{equation}
The partition function (of Laughlin edge) under dispersion $\epsilon_{m}$ reads:
\begin{equation}
    Z = \sum_{\Delta m=0}^{\infty} p(\Delta m) \cdot e^{-\beta \cdot \epsilon_{m}}.
\end{equation}
By assuming the energy dispersion is temperature independent, we can find two derivatives from this partition function:
\begin{equation}
\begin{aligned}
    \frac{\partial Z}{\partial \beta} =& - \sum_{\Delta m} p(\Delta m) \cdot e^{-\beta \epsilon_{m}} \cdot \epsilon_{m}\\
    \frac{\partial Z}{\partial (\Delta m)} =& -\beta \sum_{\Delta m} p(\Delta m) \cdot e^{-\beta \epsilon_{m}} \cdot \frac{\partial \epsilon_{m}}{\partial (\Delta m)}  
\label{eqn:SecondDerivative}
\end{aligned}
\end{equation}

From equation \ref{eqn:SecondDerivative}, we can see that:
\begin{equation}
    -\frac{1}{\beta} \frac{\partial Z}{\partial (\Delta m)} = \frac{L}{2\pi} \sum_{\Delta m} p(\Delta m) \cdot e^{-\beta \epsilon_{m}} \cdot \frac{\partial \epsilon_{m}}{\partial k}  .
\end{equation}
Take derivative with respect to $\beta$ at both side, we obtain:
\begin{equation}
    \frac{\partial}{\partial \beta} \bigg(\frac{1}{\beta} \frac{\partial Z}{\partial (\Delta m)}\bigg) = \frac{L}{2\pi} \sum_m p(\Delta m) \cdot e^{-\beta \epsilon_{m}} \cdot \frac{\partial \epsilon_{m}}{\partial k}\cdot \epsilon_{m}.
\end{equation}
 We hence deduce that the general relation between the THC and partition function is:
 \begin{equation}
    J_Q = \frac{1}{2\pi \hbar \cdot Z} \frac{\partial}{\partial \beta}\bigg(\frac{1}{\beta}\frac{\partial Z}{\partial (\Delta m)}\bigg).
\label{eqn:GeneralHeatCurrent}
\end{equation}
And the THC reads:
\begin{equation}
    \kappa = - \frac{\beta^2}{2\pi \cdot \hbar} \frac{\partial}{\partial \beta} \bigg[ \frac{1}{Z} \frac{\partial}{\partial \beta}\bigg(\frac{1}{\beta}\frac{\partial Z}{\partial (\Delta m)}\bigg) \bigg].
\label{eqn:kappaFULL}
\end{equation}
Notice that this equation holds for any dispersion relation as long as the density of states at each $m$ sector is known.

We can do the sanity check for the linear dispersion case ($\epsilon_m = v_F \Delta m$): The term inside the big bracket in Eq.~ \ref{eqn:kappaFULL} reads:
\begin{equation}
    \frac{1}{\beta} \frac{\partial Z}{\partial (\Delta m)} = -\alpha_1 \sum_{\Delta m}p(\Delta m) e^{-\beta \alpha_1 \Delta m} = - \alpha_1 Z,
\end{equation}
which gives:
\begin{equation}
    \kappa = \frac{\alpha _1 \beta^2}{2\pi \hbar} \frac{\partial}{\partial \beta} \bigg( \frac{1}{Z} \frac{\partial Z}{\partial \beta} \bigg) = \frac{\alpha _1 \beta^2}{2\pi \hbar} \frac{\partial}{\partial \beta} \bigg( \frac{\partial \log Z}{\partial \beta} \bigg) = \frac{\alpha_1 \beta^2}{h} \frac{\partial^2 \log Z}{\partial \beta^2}.
\end{equation}
Recall that $\alpha_1 = h v_F/L$, and heat capacity $C = \beta^2 \partial_{\beta}^2 \log Z$, we thus have the relation of $\kappa = \frac{v_F}{L}C$ for the linear dispersion case.

\section{Experiment Data\label{AppendixF}}
In this section, we compare the experimental error bars of the THC measured in GaAs systems (Ref.\cite{banerjee2017observed,banerjee2018observation,melcer2022absent}) with the finite-size correction term derived in this work. The correction term contains the dimensionless factor $\beta \alpha = \frac{h v_F}{k_B L T}$, which depends on three experimentally accessible parameters: the Fermi velocity $v_F$, the system size $L$, and the electron temperature $T$.

We extract the Fermi velocity from Ref.\cite{roosli2020observation,sahasrabudhe2018optimization}, where the measured value of $v_F$ for bosonic edge modes is approximately $10^5 m/s$. For the travel distant of the edge modes, we use $L=464 \mu m$ from Ref.\cite{banerjee2017observed,banerjee2018observation}, and $L =325 \mu m$ from Ref.\cite{melcer2022absent}. The THC measurements for different filling factors containing bosonic modes in Ref.\cite{banerjee2017observed,banerjee2018observation,melcer2022absent} were conducted at various temperatures, as summarized in Table \ref{table:THC_GaAs_Data}. Using these experimental parameters, we find that the corresponding values of $\beta \alpha_1$ fall within the range $0.34 \sim 1.03$. Finally, by substituting these values into our finite-size correction term, $\frac{3}{2\pi^2} \beta \alpha_1$ we estimate the expected magnitude of the correction, which provides a theoretical benchmark for interpreting the experimental error bars due to the finite-size effect of the measured THC.
\begin{table}[h]
\centering 
\renewcommand{\arraystretch}{1.6}
\begin{tabular}{c|c|c|c|c}
\hline\hline
\textbf{Filling Factor $\nu$} & \textbf{THC (in unit of $\kappa_0T$)} & \textbf{Temperature $T$} & \textbf{Estimated Value $\frac{3}{2\pi^2}\beta \alpha_1$} & \textbf{Ref.} \\ \hline
        $1$ & $0.90 \pm 0.1$ & $27\:mK$ & 0.058 & \cite{banerjee2017observed}\\ \hline
        $2$ & $0.98 \pm 0.03$ (per mode) & $30\:mK$ & 0.052 & \cite{banerjee2017observed}\\ \hline  $\frac{1}{3}$ &  $1.00 \pm 0.045$  &  $10\:mK$ & 0.160 & \cite{banerjee2017observed}\\  \hline 
         $\frac{5}{2}$(bosonic part) &  $0.97 \pm 0.03$  &  $18\:mK$ & 0.087 & \cite{banerjee2018observation}\\ \hline
         $\frac{7}{3}$&  $0.99 \pm 0.03$ (per mode)  &  $11.5\:mK$ & 0.140 & \cite{banerjee2018observation}\\ \hline
        $2$ &  $1.03 \pm 0.04$ (per mode) &  $20\: mK$ & 0.079&\cite{banerjee2018observation}\\
        \hline $1$ &  $0.93 \pm 0.03$  &  $25\: mK$ & 0.090 &\cite{melcer2022absent}
        \\ \hline
\end{tabular}
\caption{Comparison between the experimentally measured thermal Hall conductance (THC) and the estimated finite-size correction term for various filling factors $\nu$. The estimated correction term $\frac{3}{2\pi^2}\beta\alpha_1$ is computed using the experimental parameters $v_F$, $L$, and $T$ extracted from Ref.\cite{banerjee2017observed,banerjee2018observation,melcer2022absent}.}
\label{table:THC_GaAs_Data}
\end{table}


\end{document}